%% file: 00_main.tex
\pdfoutput=1
\documentclass[10pt]{article} % For LaTeX2e
\usepackage[preprint]{tmlr}
\include{01_decl}
\include{definitions}

\title{High Fidelity Neural Audio Compression}
\author{\name Alexandre D\'efossez\thanks{,$^{\dagger}$Equal contribution.} \email defossez@meta.com \\
      \addr Meta AI, FAIR Team, Paris, France
      \AND
      \name Jade Copet$^{*}$ \email jadecopet@meta.com \\
      \addr Meta AI, FAIR Team, Paris, France
      \AND
      \name Gabriel Synnaeve$^{\dagger}$ \email gab@meta.com\\
      \addr Meta AI, FAIR Team, Paris, France
      \AND
      \name Yossi Adi$^{\dagger}$ \email adiyoss@meta.com \\
      \addr Meta AI, FAIR Team, Tel-Aviv, Israel
}
\def\month{mm}  % Insert correct month for camera-ready version
\def\year{yyyy} % Insert correct year for camera-ready version
\def\openreview{\url{https://openreview.net/forum?id=??????}} % Insert correct link to

\begin{document}

\maketitle

\begin{abstract}
We introduce a state-of-the-art real-time, high-fidelity, audio codec leveraging neural networks. It consists in a streaming encoder-decoder architecture with quantized latent space trained in an end-to-end fashion.
We simplify and speed-up the training by using a single multiscale spectrogram adversary that efficiently reduces artifacts and produce high-quality samples.
We introduce a novel loss balancer mechanism to stabilize training: the \emph{weight} of a loss
now defines the fraction of the overall gradient it should represent, thus decoupling the choice of this
hyper-parameter from the typical scale of the loss. Finally, we study how lightweight Transformer models
can be used to further compress the obtained representation by up to 40\%, while staying faster than real time.
We provide a detailed description of the key design choices of the proposed model including: training objective, architectural changes and a study of various perceptual loss functions. We present an extensive subjective evaluation (MUSHRA tests) together with an ablation study for a range of bandwidths and audio domains, including speech, noisy-reverberant speech, and music. Our approach is superior to the baselines methods across all evaluated settings, considering both 24 kHz monophonic and 48 kHz stereophonic audio. Code and models are available at \href{https://github.com/facebookresearch/encodec}{github.com/facebookresearch/encodec}.

\end{abstract}

\input{01_intro}
\input{05_related.work}
\input{03_model}

\input{04_experiments}

\input{06_conclusion}
\clearpage

\bibliographystyle{tmlr}
\bibliography{refs}

\clearpage
\input{07_appendix}

\end{document}

%% file: 01_decl.tex
\usepackage[utf8]{inputenc} % allow utf-8 input
\usepackage[T1]{fontenc}    % use 8-bit T1 fonts
\usepackage{hyperref}       % hyperlinks
\usepackage{url}            % simple URL typesetting
\usepackage{booktabs}       % professional-quality tables
\usepackage{amsfonts}       % blackboard math symbols
\usepackage{nicefrac}       % compact symbols for 1/2, etc.
\usepackage{microtype}      % microtypography
\usepackage{xcolor}         % colors

\usepackage{tikz,pgfplots}
\pgfplotsset{compat=newest}
\pgfplotsset{/pgfplots/error bars/error bar style={very thick}}

\usepackage{multirow}
\usepackage{graphicx}
\usepackage{caption}
\usepackage{subcaption}
\usepackage{pifont}
\usepackage{wrapfig}
\newcommand{\pmr}[1]{\scriptsize$\pm$#1}

\usepackage{etoolbox}
\newtoggle{release}
\togglefalse{release}
\iftoggle{release}{
\newcommand{\alex}[1]{}
\newcommand{\gab}[1]{}
\newcommand{\adios}[1]{}
\newcommand{\jade}[1]{}
}
{
\newcommand{\alex}[1]{{\color{blue} A: #1}}
\newcommand{\gab}[1]{{\color{red} G: #1}}
\newcommand{\adios}[1]{{\color{magenta}Y: #1}}
\newcommand{\jade}[1]{{\color{cyan}J: #1}}

}

%% file: definitions.tex
\usepackage{bm}
\usepackage{amssymb,amsmath,graphicx,bbm,color,booktabs}
\usepackage{amsopn}

\newcommand{\vi}{\bm{i}}

\newcommand{\vx}{\bm{x}}       \newcommand{\vxh}{\hat{\bm{x}}}        
               
\newcommand{\vz}{\bm{z}}               

\newcommand{\mc}{\bm{C}}

  \newcommand{\Sc}{\mathcal{S}}

\newcommand{\encodec}{\textsc{EnCodec}~}
\renewcommand{\eqref}[1]{Eq.~(\ref{#1})}

%% file: 01_intro.tex
\section{Introduction}
%%% introduction makes claims, The body of the paper provides evidence to support each claim (forward reference it from intro)

Recent studies suggest that streaming audio and video have accounted for the majority of the internet traffic in 2021 (82\% according to~\citep{internet_traffic_cisco}).
With the internet traffic expected to grow, audio compression is an increasingly important problem. 
In lossy signal compression we aim at minimizing the bitrate of a sample while also minimizing the amount of distortion according to a given metric, ideally correlated with human perception.
Audio codecs typically employ a carefully engineered pipeline combining an encoder and a decoder 
to remove redundancies in the audio content and yield a compact bitstream. 
Traditionally, this is achieved by decomposing the input with a signal processing transform and trading off the quality of the components that are less likely to influence perception.
Leveraging neural networks as trained transforms via an encoder-decoder mechanism has been explored by~\citet{morishima1990speech,rippel2019learned,zeghidour2021soundstream}. Our research work is in the continuity of this line of work, with a focus on audio signals.

The problems arising in lossy neural compression models are twofold: first, the model has to represent a wide range of signals, such as not to overfit the training set or produce artifact laden audio outside its comfort zone. 
We solve this by having a large and diverse training set (described in Section~\ref{experiments:dataset}), as well as discriminator networks (see Section~\ref{model:losses}) that serve as perceptual losses, which we study extensively in Section~\ref{results:comparison}, Table~\ref{tab:discriminators}.
The other problem is that of compressing efficiently, both in compute time and in size. For the former, we limit ourselves to models that run in real-time on a single CPU core. For the latter, we use residual vector quantization of the neural encoder floating-point output, for which various approaches have been proposed~\citep{van2017neural,zeghidour2021soundstream}.
%have to quantize their output to get a competitive bitrate.
% Indeed, as neural networks are differentiable, their internal representations are floating point numbers. 
% Various approaches have been proposed for quantization \cite{van2017neural}. % GS STE 

% In this paper, we extend a simple differentiable quantization (DiffQ~\cite{defossez2021differentiable}) technique in Section~\ref{model:diffq}. We compare all relevant quantization methods in Section~\ref{results:comparison} (Fig.~\ref{fig:quantizers}), and show that DiffQ is state of the art across a wide range of bandwidths. \jade{This must be changed}

\begin{figure}[t!]
     \centering
     \includegraphics[trim={4cm 0 0 0},clip,width=\textwidth]{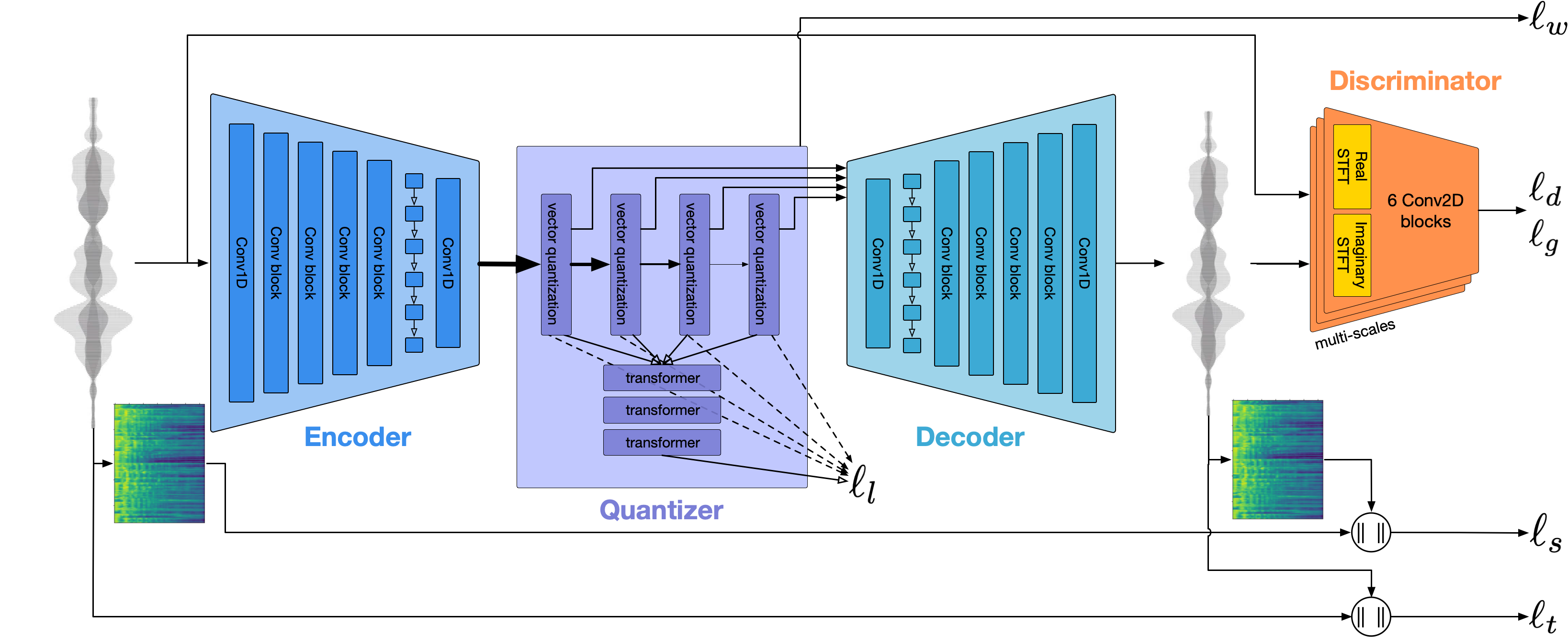}
     \caption{$\encodec$: an encoder decoder codec architecture  which is trained with reconstruction ($\ell_f$ and $\ell_t$) as well as adversarial losses ($\ell_g$ for the generator and $\ell_d$ for the discriminator). The residual vector quantization commitment loss ($\ell_w$) applies only to the encoder. Optionally, we train a small Transformer language model for entropy coding over the quantized units with $\ell_l$, which reduces bandwidth even further.
     \label{fig:arch}}
\end{figure}

Accompanying those technical contributions, we posit that designing end-to-end neural compression models is a set of intertwined choices, among which at least the encoder-decoder architecture, the quantization method, and the perceptual loss play key parts. 
Objective evaluations exist and we report scores on them in our ablations (Section~\ref{results:comparison}). But the evaluation of lossy audio codecs necessarily relies on human perception, so we ran extensive human evaluation for multiple points in this design space, both for speech and music. 
Those evaluations (MUSHRA) consist in having humans listen to, compare, and rate excerpts of speech or music compressed with competitive codecs and variants of our method, and the uncompressed ground truth.
This allows to compare variants of the whole pipeline in isolation, as well as their combined effect, in Section~\ref{results:comparison} (Figure~\ref{fig:mushra_bandwidth} and Table~\ref{tab:main_table}). 
Finally, our best model, $\encodec$, %combining our contributions, 
reaches state-of-the-art scores for speech and for music at 1.5, 3, 6, 12 kbps at 24 kHz, and at 6, 12, and 24 kbps for 48 kHz with stereo channels.

%Our main contributions are: \gab{maybe we remove this}
%\begin{itemize}
%    \item 
%\end{itemize}

%% file: 05_related.work.tex
\section{Related Work}

{\noindent \bf Speech and Audio Synthesis.} Recent advancements in neural audio generation enabled computers to efficiently generate natural sounding audio. 
The first convincing results were achieved by autoregressive models such as WaveNet~\citep{oord2016wavenet}, at the cost of slow inference.
While many other approaches were explored~\citep{parallelwavegan,wavernn,goel2022s}, the most relevant ones here are those based on Generative Adversarial Networks (GAN)~\citep{melgan, parallelwavegan, hifigan, hifi++} were able to match the quality of autoregressive  by combining various adversarial networks operate at different multi-scale and multi-period resolutions. Our work uses and extends similar adversarial losses to limit artifacts during audio generation.

% and SampleRNN~\cite{mehri2016samplernn} generate high-quality audio in the waveform domain, one sample at a time, resulting in slow inference. 
% WaveRNN~\cite{wavernn} enabled fast synthesis by training a compact neural network further optimized via sparsification. Another line of auto-regressive model involves in~\cite{goel2022s}, in the authors propose an audio generative model based on the Structured State Space Sequence model~\cite{gu2021efficiently}. Feed-forward networks were additionally suggested to further speed up inference, via knowledge distillation from an autoregressive teacher-model~\cite{parallelwavenet,ping2018clarinet, kim19flowavenet}. Two other lines of related work concerns with Flow base neural vocoders~\cite{waveglow} and diffusion based neural vocoders~\cite{wavegrad, diffwave}. Recently, generative adversarial networks (GANs)~\cite{melgan, parallelwavegan, hifigan, hifi++} were able to match the quality of autoregressive and large feed-forward models by combining various adversarial networks operate at different multi-scale and multi-period resolutions. 

{\noindent \bf Audio Codec.} Low bitrate parametric speech and audio codecs have long been studied~\citep{atal1971speech, juang1982multiple}, but their quality has been severely limited. Despite some advances~\citep{griffin1985new, mccree19962}, modeling the excitation signal has remained a challenging task. The current state-of-the-art traditional audio codecs are Opus~\citep{valin2012definition} and Enhanced Voice Service (EVS)~\citep{dietz2015overview}. These methods produce high coding efficiency for general audio while supporting various bitrates, sampling rates, and real-time compression. 

\looseness=-1
Neural based audio codecs have been recently proposed and demonstrated promising results~\citep{kleijn2018wavenet,valin2019real, lim2020robust, kleijn2021generative, zeghidour2021soundstream, omran2022disentangling, lin2022speech, jayashankar2022architecture, livariable, jiang2022end}, where most methods are based on quantizing the latent space before feeding it to the decoder. In~\citet{valin2019real}, an LPCNet~\citep{valin2019lpcnet} vocoder was conditioned on hand-crafted features and a uniform quantizer. \citet{garbacea2019low} conditioned a WaveNet based model on discrete units obtained from a VQ-VAE~\citep{van2017neural, razavi2019generating} model, while \citet{skoglund2019improving} tried feeding the Opus codec~\citep{valin2012definition} to a WaveNet to further improve its perceptual quality. \citet{jayashankar2022architecture, jiang2022end} propose an auto-encoder with a vector quantization layer applied over the latent representation and minimizing the reconstruction loss, while \citet{livariable} suggested using Gumbel-Softmax (GS)~\citep{jang2016categorical} for representation quantization. The most relevant related work to ours is the SoundStream model~\citep{zeghidour2021soundstream}, in which the authors propose a fully convolutional encoder decoder architecture with a Residual Vector Quantization (RVQ)~\citep{gray1984vector, vasuki2006review} layers. The model was optimized using both reconstruction loss and adversarial perceptual losses.

{\noindent \bf Audio Discretization.}
Representing audio and speech using discrete values was proposed to various tasks recently. \citet{dieleman2018challenge, dhariwal2020jukebox} proposed a hierarchical VQ-VAE based model for learning discrete representation of raw audio, next combined with an auto-regressive model, demonstrating the ability to generate high quality music. Similarly, \citet{lakhotia2021generative, kharitonov2021text} demonstrated that self-supervised learning methods for speech (e.g., HuBERT~\citep{hsu2021hubert}), can be quantized and used for conditional and unconditional speech generation. Similar methods were applied to speech resynthesis~\citep{polyak2021speech}, speech emotion conversion~\citep{kreuk2021textless}, spoken dialog system~\citep{nguyen2022generative}, and speech-to-speech translation~\citep{lee2021direct,lee2021textless,popuri2022enhanced}.

%% file: 03_model.tex
\section{Model}
An audio signal of duration $d$ can be represented by a sequence $\vx\in [-1, 1]^{C_\mathrm{a}\times T}$ with $C_\mathrm{a}$ the number of audio channels, $T = d\cdot f_\mathrm{sr}$  the number of audio samples at a
given sample rate $f_\mathrm{sr}$.
The $\encodec$model is composed of three main components: (i) First, an encoder network $E$ is input an audio extract and outputs a latent representation $\vz$; (ii) Next, a quantization layer $Q$ produces a compressed representation $\vz_q$,  using vector quantization;
% Notice, one can explore various quantization methods including: scalar quantization, vector quantization, residual vector quantization, etc. 
(iii) Lastly, a decoder network $G$ reconstructs the time-domain signal, $\vxh$, from the compressed latent representation $\vz_q$.
The whole system is trained end-to-end to minimize a reconstruction loss applied over both time and frequency domain, together with a perceptual loss in the form of discriminators operating at different resolutions. A visual description of the proposed method can be seen in Figure~\ref{fig:arch}. 
% distance metric $d$ (e.g., L1, L2, etc.), over either the time-domain signal~\cite{} or a frequency domain representation~\cite{}. To further improve the quality of the generated audio, recent methods suggest augmenting the objective function with various discriminators acting as a perceptual loss~\cite{}. 

\subsection{Encoder \& Decoder Architecture}
\label{model:encodec}

\looseness=-1
The $\encodec$model is a simple streaming, convolutional-based encoder-decoder architecture with sequential modeling component applied over the latent representation, both on the encoder and on the decoder side. 
Such modeling framework was shown to provide great results in various audio-related tasks, e.g., source separation and enhancement~\citep{defossez2019music, defossez2020real}, neural vocoders~\citep{melgan, hifigan}, audio codec~\citep{zeghidour2021soundstream}, and artificial bandwidth extension~\citep{tagliasacchi2020seanet, li2021real}. We use the same architecture for 24 kHz and 48 kHz audio.

% derived from the fully convolutional neural network proposed in~\cite{kumar2019melgan} and the Demucs  architecture~\cite{demusc}. Convolutional feedforward networks have been widely adopted for GAN-based vocoders, speech enhancement models and neural codecs with symmetric encoder-decoder architecture~\cite{streamingseanet,neil} or using only the decoder network as the generator~\cite{kong2020hifi,univnet}. Similarly to Demucs, we combine convolutions with auto-regressive layers in both the encoder and the decoder networks. 

{\noindent \bf Encoder-Decoder.} 
The encoder model $E$ consists in a 1D convolution with $C$ channels and a kernel size of 7 followed by $B$ convolution blocks. Each convolution block is composed of a single residual unit followed by a down-sampling layer consisting in a strided convolution, with a kernel size $K$ of twice the stride $S$. The residual unit contains two convolutions with kernel size 3 and a skip-connection. The number of channels is doubled whenever down-sampling occurred. The convolution blocks are followed by a two-layer LSTM for sequence modeling and a final 1D convolution layer with a kernel size of 7 and $D$ output channels. 
Following~\citet{zeghidour2021soundstream, li2021real}, we use $C$ = 32, $B$ = 4 and (2, 4, 5, 8) as strides. We use ELU as a non-linear activation function~\citep{clevert2015fast} either layer normalization~\citep{ba2016layer}
or weight normalization~\citep{salimans2016weight}.
We use two variants of the model, depending on whether we target the low-latency streamable setup,
or a high fidelity non-streamable usage.
With this setup, the encoder outputs 75 latent steps per second of audio at 24 kHz, and 150 at 48 kHz.
The decoder mirrors the encoder, using transposed convolutions instead of strided convolutions, and with the strides in reverse order as in the encoder, outputting the final mono or stereo audio.
%old decoder
% The decoder mirrors the encoder with $C$ channels and $B$ transposed convolution blocks. We use the same strides as the encoder in the reverse order to reconstruct the input waveform at the same resolution. First, a 1D convolution gets as input the $\vz_q$, the quantized representation. Then the architecture is composed of a two-layer LSTM followed by convolutional blocks. These blocks consist in transposed convolutions for up-sampling followed by the residual unit. The number of channels is halved whenever up-sampling occurs so that the last decoder block outputs $C$ channels. A final 1D convolution is used to output the waveform audio, either monophonic or stereophonic. 

% {\noindent \bf Non causal.} 
{\noindent \bf Non-streamable.} 
 In the non-streamable
%  non-causal 
 setup, we use for each convolution 
a total padding of $K -S$, split equally before the first time step and after the last one (with one more before if $K - S$ is odd).
We further split the input into chunks of 1 seconds, with an overlap of 10 ms to avoid clicks, and normalize each chunk before
feeding it to the model, applying the inverse operation on the output of the decoder, adding a negligible bandwidth overhead to transmit the scale.
We use layer normalization~\citep{ba2016layer}, computing the statistics including also the time dimension in order
to keep the relative scale information.

% {\noindent \bf Streaming causal.} 
{\noindent \bf Streamable.} 
For the streamable setup, all padding is put before the first time step.
For a transposed convolution with stride $s$, we output the $s$ first time steps, and keep the remaining $s$ steps in memory for completion when the next
frame is available, or discarding it at the end of a stream.
Thanks to this padding scheme, the model can output 320 samples (13 ms) as soon as the first 320 samples (13 ms)
are received. We replace the layer normalization with statistics computed over the time dimension with 
weight normalization~\citep{salimans2016weight}, as the former is ill-suited for a streaming setup.
We notice a small gain over the objective metrics by keeping a form of normalization, as demonstrated in Table~\ref{tab:arch}.

% {\noindent \bf Normalization methods.}
% In the non-streamable setup, we normalize the input audio and use LayerNorm computing the statistics over the time dimension, enabling more stable training (notice this is different than the Transformer LayerNorm which is computed over the features dimension and not over the time dimension). The causal version of this normalization strategy can be highly sensitive to spikes in the audio sequence, hence we instead opted for working on the raw audio using WeightNorm on the model as it slightly outperformed the setup without normalization method applied to the model as illustrated in Table~\ref{tab:arch}.

\subsection{Residual Vector Quantization}
\label{sec:rvq}

We use Residual Vector Quantization (RVQ) to quantize the output of the encoder as introduced by~\citet{zeghidour2021soundstream}. Vector quantization consists in projecting an input vector onto the closest
entry in a codebook of a given size. RVQ refines this process by computing the residual after quantization,
and further quantizing it using a second codebook, and so forth. 

We follow the same training procedure as described by~\citet{dhariwal2020jukebox} and \citet{zeghidour2021soundstream}. The codebook entry selected for each input is updated using an exponential moving average with a decay of 0.99, and entries that are not used are replaced with a candidate sampled from
the current batch. We use a straight-through-estimator~\citep{bengio2013estimating} to compute the gradient of the encoder, e.g. as if the quantization step was the identity function during the backward phase. Finally, a commitment loss, consisting of the MSE between the input of the quantizer and its output, with gradient only computed with respect to its input, is added to the overall training loss.

By selecting a variable number of residual steps at train time, a single model can be used to support multiple bandwidth target~\citep{zeghidour2021soundstream}. For all of our models, we use at most 32 codebooks (16 for the 48 kHz models) with 1024 entries each, e.g. 10 bits per codebook. When doing variable bandwidth training, we select randomly a number of codebooks as a multiple of 4, i.e. corresponding to a bandwidth 1.5, 3, 6, 12 or 24 kbps at 24 kHz. Given a continuous latent represention with shape $[B, D, T]$ that comes out of the encoder, this procedure turns it into a discrete set of indexes $[B, N_q, T]$ with $N_q$ the number of codebooks selected. 
This discrete representation can changed again
to a vector by summing the corresponding codebook entries, which is done just before going into the decoder.

\subsection{Language Modeling and Entropy Coding}
\label{sec:lm}

We additionally train a small Transformer based language model~\citep{vaswani2017attention} 
with the objective of 
keeping faster than real time end-to-end compression/decompression on a single CPU core. The model consists
of 5 layers, 8 heads, 200 channels, a dimension of 800 for the feed-forward blocks, and no dropout.
At train time, we select a bandwidth and the corresponding number of codebooks $N_q$. 
For a time step $t$, the discrete representation obtained at time $t - 1$ is transformed into a continuous
representation using learnt embedding tables, one for each codebook, and which are summed.
For $t=0$, a special token is used instead.
The output of the Transformer is fed into $N_q$ linear layers with as many output channels as the cardinality of each codebook (e.g. 1024), giving us the logits of the estimated distribution over each codebook for time $t$.
 We thus neglect potential mutual information between the codebooks at a single time step. This allows to speedup inference (as opposed to having one time step per codebook, or a multi-stage prediction) with a limited
impact over the final cross entropy. Each attention layer has a causal receptive field of 3.5 seconds, and we offset by a random amount the initial position of the sinusoidal position embedding to emulate being in a longer sequence. We train the model on sequences of 5 seconds.

{\noindent \bf Entropy Encoding.} 
We use a range based arithmetic coder~\citep{pasco1976source,rissanen1981universal} in order to leverage
the estimated probabilities given by the language model. As noted by~\citet{balle2018integer}, evaluation of the same model might lead to different results on different architectures, or with different evaluation procedures
due to floating point approximations. This can lead to decoding errors as the encoder and decoder will not use the exact same code. We observe in particular that the difference between batch evaluation (e.g. all time steps at once), and the real-life streaming evaluation that occurs in the decoder
can lead to difference larger than $10^{-8}$. We first round the estimated probabilities with a precision of $10^{-6}$, although evaluations in more contexts would be needed for practical deployment.
We use a total range width of $2^{24}$, and assign a minimum range width of 2. We discuss the impact on the processing time in Section~\ref{sec:latency}.

\begin{figure*}[t!]
    \centering
    \includegraphics[width=0.9\textwidth]{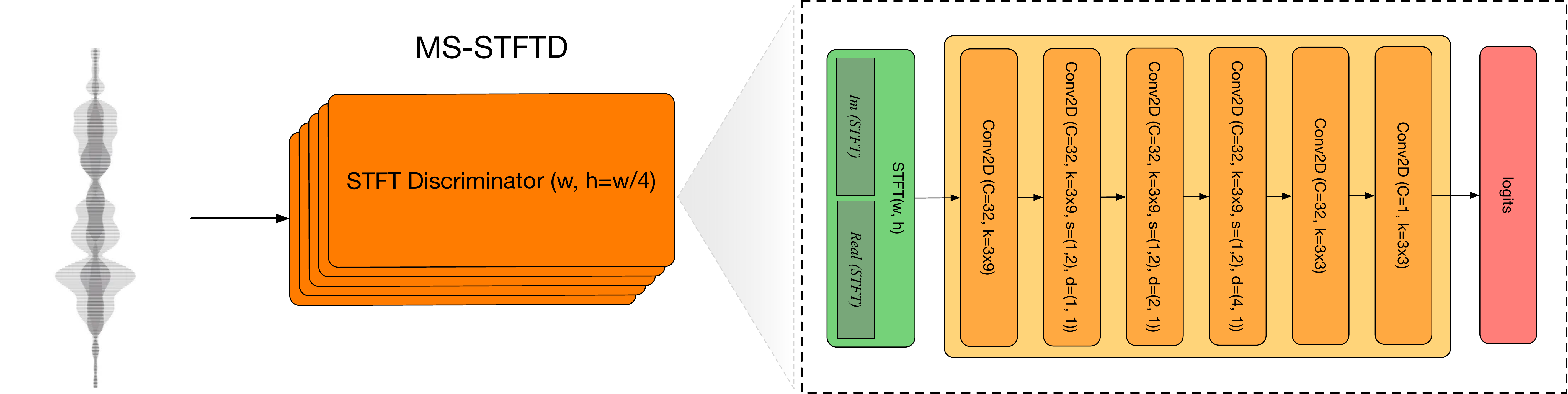}
    \caption{MS-STFT Discriminator architecture. The input to the network is a complex-valued STFT with the real and imaginary parts concatenated. Each discriminator is composed of a 2D convolutional layer, followed by 2D convolutions with increasing dilation rates. Then a final 2D convolution is applied.}
    \label{fig:dics}
\end{figure*}

\subsection{Training objective}
\label{model:losses}
We detail the training objective that combines a reconstruction loss term, a perceptual loss term (via discriminators), and the RVQ commitment loss.
% We follow a GAN based objective in which we optimize the generator and the discriminators. For the generator, we jointly optimize a reconstruction loss term, a perceptual loss term (via discriminators), and the RVQ commitment loss.
% The training objective of the generator is comprised of four loss terms: a time domain term, a frequency domain term, and two discriminator-based loss terms (adversarial loss and feature loss) acting as a perceptual loss. The discriminator loss is based on the adversarial hinge-loss function.  

{\noindent \bf Reconstruction Loss.}
The reconstruction loss term is comprised of a time and a frequency domain loss term. We minimize the L1 distance between the target and compressed audio over the time domain, i.e. $\ell_{t}(\vx, \vxh) = \|\vx - \vxh\|_1$. For the frequency domain, we use a linear combination between the L1 and L2 losses over the mel-spectrogram using several time scales~\citep{yamamoto2020parallel, gritsenko2020spectral}. 
Formally, 
\begin{equation}
    \ell_{f}(\vx, \vxh) = \frac{1}{|\alpha|\cdot |s|}\sum_{\alpha_i \in \alpha} \sum_{i \in e} \|\Sc_i(\vx) - \Sc_i(\vxh)\|_1 + \alpha_i \|\Sc_i(\vx) - \Sc_i(\vxh)\|_2,
    \label{eq:msspec}
\end{equation}
where $\Sc_i$ is a 64-bins mel-spectrogram using a normalized STFT with window size of $2^i$ and hop length of $2^i/4$, $e = {5, \ldots, 11}$ is the set of scales, and $\alpha$ represents the set of scalar coefficients balancing between the L1 and L2 terms. Unlike \citet{gritsenko2020spectral}, we take $\alpha_i = 1$.

{\noindent \bf Discriminative Loss.}
To further improve the quality of the generated samples, we introduce a perceptual loss term based on a multi-scale STFT-based (MS-STFT) discriminator, illustrated in Figure~\ref{fig:dics}.
Multi scale discriminators are popular for capturing different structures in audio signals~\citep{melgan, hifigan, you2021gan}. The MS-STFT discriminator consists in identically structured networks operating on multi-scaled complex-valued STFT with the real and imaginary parts concatenated. 
Each sub-network is composed of a 2D convolutional layer (using kernel size 3 x 8 with 32 channels), followed by 2D convolutions with increasing dilation rates in the time dimension of 1, 2 and 4, and a stride of 2 over the frequency axis. A final 2D convolution with kernel size 3 x 3 and stride (1, 1) provide the final prediction. We use 5 different scales with STFT window lengths of [2048, 1024, 512, 256, 128]. 
For 48 kHz audio, we double the size of each STFT window and train the discriminator every two batches, and for stereophonic audio, we process separately the left and right channels. We use LeakyReLU as a non-linear activation function and apply weight normalization~\citep{salimans2016weight} to our discriminator network. The MS-STFT discriminator model architecture is visually depicted in Figure~\ref{fig:dics}. 

The adversarial loss for the generator is constructed as follows, $\ell_{g}(\vxh) = \frac{1}{K} \sum_{k} \max(0, 1 - D_k(\vxh)))$, where $K$ is the number of discriminators.
Similarly to previous work on neural vocoders~\citep{melgan, hifigan, you2021gan}, we additionally include a relative feature matching loss for the generator. Formally, 
\begin{equation}
    \ell_{feat}(\vx, \vxh) = \frac{1}{KL} \sum_{k=1}^{K}\sum_{l=1}^{L} \frac{\| D_{k}^{l}(\vx) - D_{k}^{l}(\vxh) \|_1}{\mathrm{mean}\left(\|D_{k}^{l}(\vx) \|_1\right)},
    \label{eq:feat}
\end{equation}
where the $\mathrm{mean}$ is computed over all dimensions,
$(D_k)$ are the discriminators, and $L$ is the number of layers in discriminators.
The discriminators are trained to minimize the following hinge-loss adversarial loss function: $L_{d}(\vx, \vxh) = \frac{1}{K} \sum_{k=1}^{K} \max(0, 1 - D_k(\vx)) + \max(0, 1 + D_k(\vxh))$, where $K$ is the number of discriminators. Given that the discriminator tend to overpower easily the decoder, we update its weight
with a probability of 2/3 at 24 kHz, and 0.5 at 48 kHz.

{\noindent \bf Multi-bandwidth training.}
At 24 kHz, we train the model to support the bandwidths 1.5, 3, 6, 12, and 24 kbps by selecting the appropriate
number of codebooks to keep in the RVQ step, as explained in Section~\ref{sec:rvq}. At 48 kHz, we 
train to support 3, 6, 12 and 24 kbps. We also noticed that using a dedicated discriminator per-bandwidth
is beneficial to the audio quality. Thus, we select a given bandwidth for the entire batch, and evaluate
and update only the corresponding discriminator.

{\noindent \bf VQ commitment loss.}
As mentioned in Section~\ref{sec:rvq}, we add a commitment loss $l_w$ between the output of the encoder,
and its quantized value, with no gradient being computed for the quantized value.
For each residual step $c \in \{1, \ldots C\}$ (with $C$ depeding on the bandwidth target for the current batch), noting $\vz_c$ the current residual and $q_c(\vz_c)$ the nearest entry in the corresponding codebook, we define $l_w$ as
\begin{equation}
    l_w = \sum_{c=1}^{C} \| \vz_c - q_c(\vz_c)\|_2^2.
\end{equation}

Overall, the generator is trained to optimize the following loss, summed over the batch,
\begin{equation}
\label{eq:total_loss}
    \begin{split}
   &L_{G} = \lambda_t\cdot\ell_{t}(\vx, \vxh) + \lambda_f\cdot\ell_{f}(\vx, \vxh) + \lambda_g\cdot\ell_{g}(\vxh) + \lambda_{feat}\cdot\ell_{feat}(\vx, \vxh) + \lambda_w \cdot\ell_w(w),
    \end{split}
\end{equation}
% where $L_{G}$ is the generator loss, $L_{D}$ is the discriminator loss, and
where $\lambda_t$, $\lambda_f$, $\lambda_g$, $\lambda_{feat}$, and $\lambda_w$ the scalar coefficients to balance between the terms.

{\noindent \bf Balancer.}
We introduce a loss balancer in order to stabilize training, in particular the varying scale of the gradients coming from the discriminators.
We also find that the balancer makes it easier to reason about the different loss weights, independently of their scale. 
Let us take a number of losses $(\ell_i)_i$ that depends only on the output of the model $\hat{x}$.
We define $g_i = \frac{\partial \ell_i}{\partial \hat{x}}$, and $\langle \|g_i\|_2 \rangle_\beta$ the exponential moving average 
of $g_i$ over the last training batches. Given a set of weights $(\lambda_i)$ and a reference norm $R$, we 
define 
\begin{equation}
    \tilde{g}_i = R \frac{\lambda_i}{\sum_{j} \lambda_j}\cdot \frac{g_i}{\langle \|g_i\|_2 \rangle_\beta}.
\end{equation}
\looseness=-1
We then backpropagate into the network $\sum_i \tilde{g}_i$, instead of the original $\sum_i \lambda_i g_i$.
This changes the optimization problem but allows to make the $\lambda_i$ interpretable irrespectively of the natural scale of each loss.
If $\sum_i \lambda_i = 1$, then each weight can be interpreted as the fraction of the model gradient that come from the corresponding loss.
We take $R=1$ and $\beta = 0.999$. All the generator losses from \eqref{eq:total_loss} fit into the balancer, except for the commitment loss, as it is not defined with respect to the output of the model.

%% file: 04_experiments.tex
\begin{figure}[t]
     \centering
     \input{pullfigure.tikz}
     \caption{Human evaluations (MUSHRA: comparative scoring of samples) across bandwidths of standard codecs
     and neural codecs. For \encodec{} we report the 
     initial bandwidth without entropy coding (in plain) and with entropy coding (hollow). Lyra-v2 is a neural audio codec, while EVS and Opus are competitive standard codecs. The audio samples are from speech and music. The ground truth is 16bits 24kHz wave.
     \label{fig:mushra_bandwidth}}
 \end{figure}
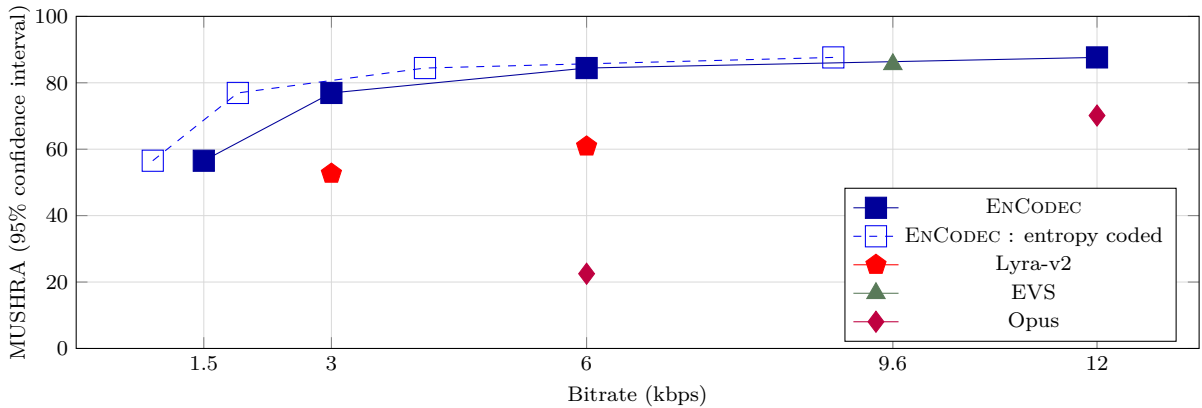

\section{Experiments and Results}

\subsection{Dataset}
\label{experiments:dataset}

\looseness=-1
We train $\encodec$ on 24 kHz monophonic across diverse domains, namely: speech, noisy speech, music and general audio while we train the fullband stereo $\encodec$ on only 48 kHz music. 
For speech, we use the clean speech segments from DNS Challenge 4~\citep{dns} and the Common Voice dataset~\citep{ardila2019common}. For general audio, we use on AudioSet~\citep{gemmeke2017audio} together with FSD50K~\citep{fonseca2021fsd50k}. 
For music, we rely on the Jamendo dataset~\citep{bogdanov2019mtg} for training and evaluation and we further evaluate our models on music using a proprietary music dataset. Data splits are detailed in Appendix~\ref{sec:exp_details}. 

For training and validation, we define a mixing strategy which consists in either sampling a single source from a dataset or performing on the fly mixing of two or three sources. Specifically, we have four strategies: (s1) we sample a single source from Jamendo with probability 0.32; (s2) we sample a single source from the other datasets with the same probability; (s3) we mix two sources from all datasets with a probability of 0.24; (s4) we mix three sources from all datasets except music with a probability of 0.12. 

The audio is normalized by file and we apply a random gain between -10 and 6 dB. We reject any sample that has been clipped. Finally we add reverberation using room impulse responses provided by the DNS challenge with probability 0.2, and RT60 in the range [0.3, 1.3] except for the single-source music samples.
For testing, we use four categories: clean speech from DNS alone, clean speech mixed with FSDK50K sample, Jamendo sample alone, proprietary music sample alone.

\subsection{Baselines}

Opus~\citep{valin2012definition} is a versatile speech and audio codec standardized by the IETF in 2012. 
It scales from 6 kbps narrowband monophonic audio to 510 kbps fullband stereophonic audio. EVS~\citep{dietz2015overview} is a codec standardized in 2014 by 3GPP and developed for Voice over LTE (VoLTE).
It supports a range of bitrates from 5.9 kbps to 128 kbps, and audio bandwidths from 4 kHz to 20 kHz. It is the successor of AMR-WB~\citep{bhagat2012adaptive}.% and retains full backwards compatibility.
% Both Opus and EVS have been widely deployed for speech communication, audiovisual conferencing services and audio streaming over the internet and serve billions of daily users.
We use both codecs to serve as traditional digital signal processing baselines. We also utilize MP3 compression at 64 kbps as an additional baseline for the stereophonic signal compression case. MP3 uses lossy data compression by approximating the accuracy of certain components of sound that are considered to be beyond hearing capabilities of most humans. Finally, we compare $\encodec$ to the SoundStream model from the official implementation available in Lyra 2~\footnote{https://github.com/google/lyra} at 3.2 kbps and 6 kbps on audio upsampled to 32 kHz.
We also reproduced a version of SoundStream~\citep{zeghidour2021soundstream} with minor improvements. Namely, we use the relative feature loss introduce in Section~\ref{model:losses}, and layer normalization (applied separately for each time step) in the discriminators, except for the first and last layer, which improved the audio quality during our preliminary studies. Results a reported in Table~\ref{tab:comp} in the Appendix~\ref{sec:add_res}.

% over signals sampled at 24 kHz~\footnote{As no official implementation is provided, we use our own implementation for the SoundStream model, after reaching comparable results to the ones reported in the SoundStream paper.}. 
% SoundStream model was developed to support for operates on signals sampled at 24 kHz and supports a range of bitrates from 3kbps to 18kbps, with the option of having a single model handling multiple bitrates. It is relying on a fully convolutional encoder-decoder architecture inspired from \cite{melgan}, a residual vector quantizer and it was trained with a combination of adversarial and reconstruction losses for audio generation.  

% \adios{rephrase this paragraph to include only diffq and our won soundstream implementation}
% We additionally introduce and evaluate another fully differentiable vector quantizer based on Gumbel-Softmax (GS)~\cite{jang2016categorical}. In which, we follow the same neural architecture and replace the quantization layer with a GS layer. A formal definition of the GS quantizer can be found in Appendix~\ref{sec:model:gs}.

% \vspace{-0.1cm}
\subsection{Evaluation Methods}
\label{eval_methods}
We consider both subjective and objective evaluation metrics. For the subjective tests we follow the MUSHRA protocol~\citep{series2014method}, using both a hidden reference and a low anchor. Annotators were recruited using a crowd-sourcing platform, in which they were asked to rate the perceptual quality of the provided samples in a range between 1 to 100. We randomly select 50 samples of 5 seconds from each category of the the test set and force at least 10 annotations per samples. To filter noisy annotations and outliers we remove annotators who rate the reference recordings less then 90 in at least 20\% of the cases, or rate the low-anchor recording above 80 more than 50\% of the time. 
For objective metrics, we use ViSQOL~\citep{hines2012visqol,chinen2020visqol}~\footnote{We compute visqol with: \url{https://github.com/google/visqol} using the recommended recipes.}, together with the Scale-Invariant Signal-to-Noise Ration (SI-SNR)~\citep{luo2019conv, nachmani2020voice, chazan2021single}. 

\subsection{Training}
\label{training}

We train all models for 300 epochs, with one epoch being 2,000 updates with the Adam optimizer with a batch size of 64 examples of 1 second each, a learning rate of $3\cdot 10^{-4}$, $\beta_1=0.5$, and $\beta_2=0.9$. 
All the models are traind using 8 A100 GPUs. We use the balancer introduced in Section~\ref{model:losses}
with weights
$\lambda_t = 0.1$, $\lambda_f = 1$, $\lambda_g=3$, $\lambda_\mathrm{feat} = 3$ for the 24 kHz models.
For the 48 kHz model, we use instead $\lambda_g=4$, $\lambda_\mathrm{feat} = 4$.

% We train on 4 A100 GPUs hosted on AWS for most experiments except the 48 kHz for which 8 GPUs are used.

\subsection{Results}
\label{results}

\begin{table}[t]
    \caption{MUSHRA scores for Opus, EVS, Lyra-v2, and $\encodec$ for various bandwidths under the streamable setting. Results are reported across different audio categories (clean speech, noisy speech, and music), sampled at 24 kHz. We report mean scores and 95\% confidence intervals.
    For \encodec{}, we also report the average bandwidth after using the entropy coding described in Section~\ref{sec:lm}. \label{tab:main_table}}
  \centering
  \setlength\tabcolsep{4pt}
  \resizebox{0.95\textwidth}{!}{
  \begin{tabular}{lrrccccccc}
    \toprule
    Model & Bandwidth & Entropy Coded & Clean Speech & Noisy Speech & Music Set-1 & Music Set-2\\
    \midrule
    Reference & - & - & 95.5\pmr{1.6} & 93.9\pmr{1.8} & 93.2\pmr{2.5} & 97.1\pmr{1.3} \\
    \midrule
    Opus & 6.0 kbps  & - &30.1\pmr{2.8} & 19.1\pmr{5.9} & 20.6\pmr{5.8} & 17.9\pmr{5.3} \\
    Opus & 12.0 kbps & - & 76.5\pmr{2.3} & 61.9\pmr{2.1} & 77.8\pmr{3.2} & 65.4\pmr{2.7} \\
    \midrule
    EVS  & 9.6 kbps & - & 84.4\pmr{2.5} & 80.0\pmr{2.4} & 89.9\pmr{2.3} & 87.7\pmr{2.3} \\
    \midrule
    Lyra-v2  & 3.0 kbps & - & 53.1\pmr{1.9} & 52.0\pmr{4.7} & 69.3\pmr{3.3} & 42.3\pmr{3.5} \\
    Lyra-v2  & 6.0 kbps & - &66.2\pmr{2.9} & 59.9\pmr{3.3} & 75.7\pmr{2.6} & 48.6\pmr{2.1} \\
    \midrule
    \encodec  & 1.5 kbps & 0.9 kbps &49.2\pmr{2.4} & 41.3\pmr{3.6} & 68.2\pmr{2.2} & 66.5\pmr{2.3} \\
    \encodec  & 3.0 kbps & 1.9 kbps& 67.0\pmr{1.5} & 62.5\pmr{2.3} & 89.6\pmr{3.1} & 87.8\pmr{2.9} \\
    \encodec  & 6.0 kbps & 4.1 kbps & 83.1\pmr{2.7} & 69.4\pmr{2.3} & 92.9\pmr{1.8} & 91.3\pmr{2.1} \\
    \encodec  & 12.0 kbps & 8.9 kbps & 90.6\pmr{2.6} & 80.1\pmr{2.5} & 91.8\pmr{2.5} & 92.9\pmr{1.2} \\
    \bottomrule
  \end{tabular}}
\end{table}

\looseness=-1
We start with the results for $\encodec$ with a bandwidth in $\{1.5, 3, 6, 12\}$ kbps and compare them to the baselines. Results for the streamable setup are reported in Figure~\ref{fig:mushra_bandwidth} and a breakdown per category in Table~\ref{tab:main_table}. We additionally explored other quantizers such as Gumbel-Softmax and DiffQ (see details in Appendix~\ref{sec:abb_quant}), however, we found in preliminary results that they provide similar or worse results, hence we do not report them.

When considering the same bandwidth, \encodec is superior to all evaluated baselines considering the MUSHRA score. Notice, \encodec at 3kbps reaches better performance on average than Lyra-v2 using 6kbps and Opus at 12kbps. When considering the additional language model over the codes, we can reduce the bandwidth by $\sim 25-40\%$. For instance, we can reduce the bandwidth of the 3 kpbs model to 1.9 kbps. We observe that for higher bandwidth, 
the compression ratio is lower, which could be explained by the small size of the Transformer model used, making
hard to model all codebooks together.

\subsubsection{Ablation study}
\label{results:comparison}
Next, we perform an ablation study to better evaluate the effect of the discriminator setup, streaming, multi-target bandwidth, and balancer. We provide more detailed ablation studies in the Appendix, Section~\ref{sec:add_res}.

{\noindent \bf The effect of discriminators setup.}
Various discriminators were proposed in prior work to improve the perceptual quality of the generated audio. The Multi-Scale Discriminator (MSD) model proposed by~\citet{melgan} and adopted in~\citep{hifigan, hifi++, zeghidour2021soundstream}, operates on the raw waveform at different resolutions. We adopt the same MSD configuration as described in~\citet{zeghidour2021soundstream}. \citet{hifigan} additionally propose the Multi-Period Discriminator (MPD) model, which reshapes the waveform to a 2D input with multiple periods.
% We set the periods to [1, 2, 3, 4, 5] on 24 kHz audio as it greatly improves the audio reconstruction quality and MOS scores compared to the original periods of [2, 3, 5, 7, 11] that appears to be better suited for 48 kHz.
Next, the STFT Discriminator (Mono-STFTD) model was introduced in~\citet{zeghidour2021soundstream}, where a single network operates over the complex-valued STFT. 

\begin{table}[t!]
  \caption{Comparing discriminators using objective (ViSQOL, SI-SNR) and subjective metrics (MUSHRA).\label{tab:discriminators}}
  \centering
  \begin{tabular}{lccccc}
    \toprule
    Discriminator setup & SI-SNR & ViSQOL & MUSHRA\\
    \midrule
    MSD+Mono-STFT         & 5.99 & 4.22 & 62.91\pmr{2.62}\\
    % MSD+Mono-STFT         & \textbf{6.67} & \textbf{4.38} & 47.2\pmr{3.4}\\
    MPD                   & 7.35 & 4.24 & 60.7\pmr{2.8}\\
    MS-STFT+MPD           & 6.55 & \textbf{4.34} & \textbf{79.0\pmr{1.9}}\\
    \midrule
    MS-STFT               & \textbf{6.67} & \textbf{4.35} & \textbf{77.5\pmr{1.8}}\\
    \bottomrule
  \end{tabular}
\end{table}

\looseness=-1
We evaluate our MS-STFTD discriminator against three other discriminator configurations: (i) MSD+Mono-STFTD (as in~\citet{zeghidour2021soundstream}); (ii) MPD only; (iii) MS-STFTD only; (vi) MS-STFTD+MPD. Results are reported in Table~\ref{tab:discriminators}. Results suggest that using only a multi-scale STFT-based discriminator such as MS-STFTD, is enough to generate high quality audio. Additionally, it simplifies the model training and reduces training time. Including the MPD discriminator, adds a small gain when considering the MUSHRA score.

\begin{table}[t!]
  \caption{Streamable vs. Non-streamable evaluations at 6 kbps on an equal mix of speech and music.\label{tab:cnc}}
  \centering
  \begin{tabular}{lcccc}
    \toprule
    Model & Streamable &  SI-SNR & ViSQOL\\
    \midrule
    Opus     & \ding{51}  & 2.45 & 2.60\\
    EVS      & \ding{51}  & 1.89 & 2.74\\
    \encodec & \ding{51}  & 6.67 & 4.35\\
    \encodec & \ding{55}  & 7.46 & 4.39\\
    \bottomrule
  \end{tabular}
\end{table}

{\noindent \bf The effect of the streamable modeling.}
We also investigate streamable vs. non-streamable setups and report results in Table~\ref{tab:cnc}. Unsurprisingly, we notice a small degradation switching from non-streamable to streamable but the performance remains strong while this setting enables streaming inference.

{\noindent \bf The effect of the balancer.}
Lastly, we present results evaluating the impact of the balancer. We train the $\encodec$ model considering various values $\lambda_t$, $\lambda_f$, $\lambda_g$, and $\lambda_{feat}$ with and without the balancer. Results are reported in Table~\ref{tab:balancer} in the Appendix. As expected, results suggest the balancer significantly stabilizes the training process. See Appendix~\ref{sec:add_res} for more details. 

\subsubsection{Stereo Evaluation}
\label{results:stereo}
\begin{table}
  \centering
  \caption{\textbf{Stereophonic} extreme music compression versus MP3 and Opus for music sampled at 48 kHz. 
  \label{tab:stereo}}
  \begin{tabular}{lrrrc}
    \toprule
    Model & Bandwidth & Entropy Coded & Compression &  MUSHRA \\
    \midrule
    Reference & - & - & 1$\times$ & \textbf{95.1\pmr{1.8}}\\
    \midrule
    MP3 & 64 kbps & - & 24$\times$            & 82.7\pmr{3.2}\\
    Opus & 6 kbps& - & 256$\times$          & 17.7\pmr{5.9}\\
    Opus & 24 kbps& - &64$\times$         & 82.9\pmr{3.7}\\
    \encodec & 6  kbps& 4.2 kbps & 256$\times$     & 82.9\pmr{2.4} \\
    \encodec & 12 kbps& 8.9 kbps & 128$\times$	 & \textbf{88.0\pmr{2.7}} \\
    \encodec & 24 kbps& 19.4 kbps & 64$\times$	 & \textbf{87.5\pmr{2.6}} \\
    \bottomrule
    \end{tabular}
\end{table}

All previously reported results considered only the monophonic setup. Although it makes sense when considering speech data, however for music data, stereo compression is highly important. We adjust our current setup to stereo by only modifying our discriminator setup as described in Section~\ref{model:losses}.

Results for $\encodec$ working at 6 kbps, $\encodec$ with Residual Vector Quantization (RVQ) at 6 kbps, and Opus at 6 kbps, and MP3 at 64 kbps are reported in Table~\ref{tab:stereo}. \encodec is significantly outperforms Opus at 6kbps and is comparable to MP3 at 64kbps, while \encodec at 12kpbs achieve comparable performance to \encodec at 24kbps. Using a language model and entropy coding gives a variable gain
between 20\% to 30\%.

\subsection{Latency and computation time}
\label{sec:latency}

\begin{table}
  \centering
  \caption{Initial latency and real time factor (RTF) for Lyra~v2, \encodec at 24 kHz and 48 kHz. A RTF greater than 1 indicates faster than real time processing.
  We report the RTF for both the encoding (Enc.) and decoding (Dec.), without and 
  with entropy coding (EC). All models are evaluated at 6 kbps.
  \label{tab:latency}}
    % \resizebox{0.8\textwidth}{!}{
  \begin{tabular}{lcrrrr}
    \toprule
    & & \multicolumn{4}{c}{Real Time Factor}\\
    \cmidrule{3-6}
    Model & Latency & \emph{Enc.} & \emph{Dec.} &  \emph{Enc. + EC} & \emph{Dec. + EC}  \\
    \midrule
    Lyra v2 (32 kHz) & - & 27.4 & 67.2 & - & - \\
    \midrule
    \encodec 24 kHz & 13 ms & 9.8 & 10.4 & 1.6 & 1.6\\
    \encodec 48 kHz & 1 s & 6.8 & 5.1 & 0.68 & 0.66\\
    \bottomrule
    \end{tabular}
    % }
\end{table}

We report the initial latency and real time factor on Table~\ref{tab:latency}.
The real-time factor is here defined as the ratio between the duration of the audio and the processing time, so that it is greater than one when the method is faster than real time. We profiled all models on a single thread of a MacBook Pro 2019 CPU at 6 kbps. 

\looseness=-1
{\noindent \bf Initial latency.}
The 24 kHz streaming \encodec{} model has an initial latency (i.e., without the computation time) of 13.3 ms.
The 48 kHz non-streaming version has an initial latency of 1 second, due to the normalizations used.
Note that using entropy coding increases the initial latency, because the stream
cannot be ``flushed'' with each frame, in order to keep the overhead small. 
Thus decoding the frame at time $t$, requires for the frame $t + 1$ to be partially received, increasing the latency by 13ms.

{\noindent \bf Real time factor.}
While our model is worse than Lyra~v2 in term of processing speed, it processes the audio 10 times
faster than real time, making it a good candidate for real life applications. The gain from the entropy coding comes at a cost, although the processing is still faster than real time and could be used
for applications where latency is not essential (e.g. streaming). At 48 kHz, the increased number of step size lead to a slower than real time processing, although a more efficient implementation, or using 
accelerated hardware would improve the RTF. It could also be used for archiving where real time processing is not required.

%% file: pullfigure.tikz
\begin{tikzpicture}
\begin{footnotesize}
\begin{axis}[
  xlabel=Bitrate (kbps),
  ylabel=MUSHRA (95\% confidence interval),
  ylabel style={yshift=-0.5em},
  ymax=100,
  ymin=0,
  xmin=0,
  xtick={1.5, 3, 6, 9.6, 12, 24},
  xticklabels={1.5, 3, 6, 9.6, 12, 384},
  %xmode=log,
  %log ticks with fixed point,
  %ymode=log,
  legend style={at={(0.98,0.02)},anchor=south east},
  width=\columnwidth,
  height=6.0cm,
  error bars/y dir=both,
  error bars/y explicit=true,
  grid style={line width=.1pt, draw=gray!30},
  grid=both,
]
\addplot+ [color=black!40!blue!100, mark=square*, mark options={scale=2}] table [
                x=Bitrate,
                y=Encodec, 
                %y error=EncodecError] 
                ]{pullfigure.dat};
\addlegendentry{\encodec}
\addplot+ [color=black!05!blue!100, mark=square, mark options={scale=2, solid}, solid, dashed] table [
                y=EntropyCoded, 
                %y error=EntropyCodedError] 
                ]{pullfigure.dat};
\addlegendentry{\encodec: entropy coded}
\addplot+ [solid, color=red, mark=pentagon*, mark options={scale=2}] table [scatter,only marks, y=Lyrav2, x=Bitrate, 
%y error=Lyrav2Error,
]{pullfigure.dat};
\addlegendentry{Lyra-v2}
\addplot+ [solid, color=black!80!green!65, mark=triangle*, mark options={scale=2}] table [scatter,only marks, y=EVS, x=Bitrate, 
%y error=EVSError,
]{pullfigure.dat};
\addlegendentry{EVS}
\addplot+ [solid, color=purple, mark=diamond*, mark options={scale=2}] table [scatter,only marks, y=Opus, x=Bitrate, 
%y error=OpusError,
]{pullfigure.dat};
\addlegendentry{Opus}
%\addplot+ [solid, color=olive, mark=+, mark options={scale=1.5}] table [scatter,only marks, y=GT, x=Bitrate, y error=GTError,
]{pullfigure.dat};
%\addlegendentry{Ground truth}

\end{axis}
\end{footnotesize}
\end{tikzpicture}

%% file: 06_conclusion.tex
\vspace{-0.1cm}
\section{Conclusion}
\label{sec:con}
\vspace{-0.1cm}
We presented $\encodec$: a state-of-the-art real-time neural audio compression model, producing high-fidelity audio samples across a range of sample rates and bandwidth. We showed subjective and objective results from 24kHz monophonic at 1.5 kbps (Figure~\ref{fig:mushra_bandwidth}) to 48kHz stereophonic (Table~\ref{tab:stereo}). 
We improved sample quality by developing a simple but potent spectrogram-only adversarial loss which efficiently reduces artifacts and produce high-quality samples. 
Besides, we stabilized training and improved the interpretability of the weights for losses through a novel gradient balancer. 
Finally, we also demonstrated that a small Transformer model can be used to further reduce the bandwidth by up to 40\% without further
degradation of quality, in particular for applications where low latency is not essential (e.g. music streaming).

%% file: 07_appendix.tex
\appendix

\counterwithin{figure}{section}
\counterwithin{table}{section}

\section{Appendix}

\subsection{Experimental details}
\label{sec:exp_details}

{\noindent \bf Datasets details.} 
We present additional details and statistics over the datasets used for training in Table~\ref{tab:datasets}. Notice that for some datasets, each sub-source or sample has its own license and we then refer the reader to the respective dataset details for more information. 
We create our datasets splits as followed. For Common Voice, we randomly sample 99.5\% of the dataset for train, 0.25\% for valid and the rest for test splits. Similarly, we sample 98\% of the clean segments from DNS Challenge 4 for train, 1\% for valid and 1\% for test.
For AudioSet, we use the unbalanced train segments as training data and randomly selected half of the eval segments as validation set and the other half as test set. We follow the same procedure for FSD50K using the dev set for training and splitting the eval set between validation and test. Finally for the Jamendo dataset, we randomly take 96\% of the artists and their corresponding tracks for train, 2\% for valid and 2\% for test, hence there is no artists overlap in the different sets.

{\noindent \bf SoundStream model.}
We additionally re-implemented SoundStream. We follow the implementation details in ~\citet{zeghidour2021soundstream} to develop our own SoundStream version as the original implementation is not open sourced.  
We implement Residual Vector Quantization with k-means initialization, exponential moving average updates and random restart as pointed by the original implementation.
For the wave-based discriminator, we follow the details provided in ~\citet{zeghidour2021soundstream} and refer to the original Multi-Scale Discriminator implementation proposed in~\citet{melgan} for additional information. We use LeakyReLU as non-linear activation function and following \citet{hifigan}, we use spectral normalization for the original resolution and weight normalization for other resolutions. For the STFT-based discriminator, we experimented with multiple normalization methods and found that using LayerNorm~\citep{ba2016layer} was the only one that prevented the discriminator from diverging. We used LeakyReLU as non-linear activation function.
Finally, training hyper parameters are not shared either so we use the same parameters as for our $\encodec$ model.  

\begin{table}[t!]
  \caption{Datasets description. License with asterisk annotation * imply that the specific license varies across the dataset and is specific to each sample.
  \label{tab:datasets}}
  \centering
  \resizebox{0.9\textwidth}{!}{
  \begin{tabular}{llrrrl}
    \toprule
    Dataset & Audio domain & Sampling rate & Channels & Duration & License\\
    \midrule
    Common Voice 7.0 & Speech & 48 kHz & 1 & 9,096 h & CC-0\\
    DNS Challenge 4 (speech) & Speech & 48 kHz & 1 & 2,425 h & Multiples*\\
    AudioSet & General audio & 48 kHz & 2 & 4,989 h & CC BY 4.0*\\
    FSD50K & General audio & 44.1 kHz & 1 & 108 h & CC*\\
    Jamendo & Music & multiples & 2 & 919 h & CC*\\
    \bottomrule
  \end{tabular}}
\end{table}

\subsection{Alternative quantizers}
\label{sec:abb_quant}

\subsubsection{DiffQ Quantizer}
\label{model:diffq}
\textbf{Pseudo quantization noise.}
We perform scalar quantization of the latent representation. As quantization is non differentiable,
we use properly scaled additive independent noise, a.k.a pseudo quantization noise, at train time to simulate it. This approach was first use for analog-to-digital converters design~\citep{widrow1996statistical}, then for image compression~\citep{balle2017end},
and finally by the DiffQ model compression method~\citep{defossez2021differentiable} with a differentiable bandwidth estimate. 
We extend the DiffQ approach for latent space quantization, adding
support for streamable 
% causal 
rescaling, proper sparsity, and improved prior coding. Formally, we introduce 
a learnt parameter $B\in \mathbb{R}^D$ (with $D$ the dimension of the latent space) such that $B^{(i)}$ represents the number of bits to use of the $i$-th dimension. 
In practice $B$ is parameterized as $B = B_{\max} \cdot \mathrm{sigmoid}(\alpha v)$, with $B_{\max} = 15$, and the $\alpha = 5$ factor is used for boosting the learning rate of $v$, the learnt parameter. We then define the pseudo quantized representation $\vz_{q,\mathrm{train}}$ used at train time,
\begin{equation}
    \label{eq:dq_train}
    \vz_{q,\mathrm{train}} = \mathrm{clamp}(\vz, m - L\cdot\sigma, m + L\cdot\sigma) + L\cdot\sigma\cdot  \frac{\mathcal{U}[-1, 1]}{2^B},
\end{equation}
with $m$ (resp. $\sigma$) the mean (resp. standard deviation) of $z$ along the time and batch axis, and $\mathcal{U}[-1, 1]$ uniform i.i.d noise.
The limit $L$ is chosen so that when $B$ goes to 0, the noise covers most of the range of values accessible to a Gaussian variable of variance $\sigma$.
In order to prevent outlier values, we clamp the input $\vz$ to this expected range of values. If $L$ is too large,
the dynamic range of the quantization will be poorly used. If $L$ is too low, many values will get clamped and lose gradients. In practice we choose $L = 3$, which verifies that values are not clamped 99.5\% of the time (against 99.8\% if the input were gaussian).
In order to learn $B$, we approximate at train time the bandwidth used by the model by $w_\mathrm{diffq} = T' \sum_{i=1}^D B^{(i)} / d$ with $T'$ the number of latent time steps, $d$ the duration of the sample,
and add to the training loss a penalty term of the form $\lambda w_\mathrm{diffq}$, as long as $w_\mathrm{diffq}$ is over a given target, as used in Section~\ref{model:losses}.

\textbf{Test time quantization. }At validation and test time, we replace the batch-level statistic with an exponential moving average with a decay of $0.9$ computed over the train set,
similar to batch norm~\citep{batchnorm}.
We first normalize and clamp $\vz$ to the segment $[0, 1]$ as $u = \mathrm{clamp}((L + \sigma^{-1} (\vz - m)) / 2L, 0 , 1)$ and define the number of quantized
levels $N_B = \mathrm{round}(2^B)$ and the quantized index as
\begin{equation}
    \label{eq:quant_index}
    \vi = \min\left[\mathrm{floor}(N_B \cdot u), N_B - 1\right] \in [0..N_B - 1],
\end{equation}
with the minimum taken to avoid the edge case $u = 1$. We know we can code each entry in $\vi$ on at most $\log_2(N_B)$ bits.
The quantized latent $\vz_q$ is finally defined as
    $\vz_q = m + L \sigma \left(2 \frac{\vi + 0.5}{N_B} - 1\right)$.
% Remember that $m$ and $\sigma$ are frozen into the model as the exponential moving average result, so that the method is causal at test time.

\textbf{Sparsity. }
We want to allow $B$ to go to 0, however additive noise fails to remove all information contained in the latent in that case.
Indeed, if $\vz < m$ for instance, then even after adding the largest possible amount of noise, $\vz_{q, \mathrm{train}}$ is still biased
towards values smaller than $m$. In order to remove all information about $\vz$ from $\vz_{q,\mathrm{train}}$, the scale of the noise
relative to the scale of the signal must go to infinity. This is achieved by scaling down $\vz$ by a factor $\min(B, 1)$
in \eqref{eq:dq_train}, while scaling down the additive noise only by a factor $\sqrt{\min(B, 1)}$. Thus, the decoder
cannot invert the downscaling of $\vz$ without blowing up the noise. In the limit of $B \rightarrow 0$, we recover a sparse representation.

% \textbf{Improved prior and multi-bandwidth.}
% The quantized code can easily be coded over $\log_2(N_B)$ bits, assuming a uniform prior. A small gain is obtained by instead using a discretized gaussian prior 
% of mean $m$ and variance $\sigma$ at test time and using arithmetic coding.
% Arithmetic coding is convenient for the streaming setup as incomplete bytes from coding one latent time step can be be used to code the next one, allowing
% for near optimal coding at the code of one overall stride of latency. When using this prior, there is a mismatch between the train and test bandwidth. We account
% for that by using the $\lambda w_{\mathrm{diffq}}$ term only when the effective test time bandwidth is over the target.
% Finally, we simply support multi-bandwidths quantization by training one set of $B$ parameters per target bandwidth.

% When using the DiffQ quantizer from Section~\ref{model:diffq} or the Gumbel-softmax one (Section~\ref{sec:model:gs}), we leverage the differentiable bandwidth estimator $w$ given by either method. Given a target bandwidth $w_\mathrm{target}$,
% we add a bandwidth penalty term $\ell_w(w) = \max(w, w_\mathrm{target})$.

\subsubsection{Gumbel softmax quantizer}
\label{sec:model:gs}

We introduce a second fully differentiable vector quantizer
composed of $N_C$ codebooks each with $\Omega$ entries. The $i$-th codebook is composed of a set of centroids $\mc_i\in \mathbb{R}^{\Omega\times D}$, a logit bias $b_i\in \mathbb{R}^{\Omega}$, and a learnt prior logit $l_i \in \mathbb{R}^{\Omega}$. Assuming for simplicity a latent vector for the $j$-th time step $\vz_j \in\mathbb{R}^D$, we define a probability distribution over each codebook entries as $q_i(z) = \mathrm{softmax}(\mc_i \vz_j + b_i)$. We then sample from the corresponding gumbel-softmax~\citep{jang2016categorical} with a temperature $\tau = 0.5$. This gives
us a differentiable approximately 1-hot vector over the codebooks, i.e., noting $\mathrm{GS}$ the gumbel-softmax,
\begin{equation}
    \vz_{q, \mathrm{train}} = \sum_{i=1}^{N_C} \mathrm{GS}(\log(q_i(\vz)), \tau)^T \mc_i.
\end{equation}
At test time, we replace the gumbel-softmax with a sampling from the distribution $q_i$. We define for all $i$, $p_i = \mathrm{softmax}(l_i)$ the prior
distribution over the codebooks entries which is used for coding the quantized value $\vz_q$ with an arithmetic coder. 
We can both train the prior and minimize the bandwidth with a single 
loss term given by the cross entropy between the prior $p_i$ and the posterior $q_i(\vz)$, i.e. the differentiable bandwidth estimate for this method is given by $w_\mathrm{gs} = \sum_{i=1}^{N_C}  \sum_{k=1}^{\Omega} -q_i(\vz) \log(p_i)$.

% \subsection{Model Details}
% \label{sec:model_details}
% {\noindent \bf MS-STFT Discriminator.} 
% The MS-STFT Discriminator model architecture is detailed in Figure~\ref{fig:dics}. We use LeakyReLU as a non-linear activation function and apply weight normalization~\cite{salimans2016weight} to our discriminator network.
% For our discriminator study, we use the MSD and STFT implementation specified in~\cite{zeghidour2021soundstream}, and the MPD architecture detailed in~\cite{hifigan}.
% with the difference that we set the periods to [1, 2, 3, 4, 5] on 24 kHz audio. We noticed that this set of periods greatly improves the audio reconstruction quality compared to the original periods of [2, 3, 5, 7, 11] that appears to be better suited for 48 kHz.

\subsection{Additional Results}
\label{sec:add_res}

{\noindent \bf Comparing to SoundStream.}
For fair evaluation, we also compare \encodec to our reimplementation of SoundStream~\citep{zeghidour2021soundstream}. For which, the quantizer corresponds to RVQ in and the discriminator corresponds to STFTD+MSD. Results are reported in Table~\ref{tab:comp}. The results of our SoundStream implementation are slightly worse than \encodec using DiffQ quantizer. However, when considering the RVQ as the latent quantizer, \encodec is superior to the SoundStream model. Both \encodec and SoundStream methods are significantly better than Opus and EVS at 6.0kbps.

\begin{table}[t!]
  \caption{A comparison between \encodec using either DiffQ or RVQ as latent quantizers against our implementation of the SoundStream model at 3kbps. We additionally include the results of Opus and EVS at 6.0kbps for reference.\label{tab:comp}}
  \centering
  \begin{tabular}{lcc}
    \toprule
    Model  & Bandwidth & MUSHRA \\
    \midrule
    Reference        & -   & 96.1\pmr{1.41}\\
    \midrule
    Opus             & 6.0 & 21.1\pmr{2.62}\\
    EVS              & 6.0 & 62.9\pmr{2.18}\\
    SoundStream      & 3.0 & 71.8\pmr{1.51}\\
    \encodec (DiffQ) & 3.0 & 72.3\pmr{1.18}\\
    \encodec (RVQ)   & 3.0 & \textbf{76.8\pmr{1.31}}\\
    \bottomrule
  \end{tabular}
\end{table}

{\noindent \bf The effect of the model architecture.} 
We investigate the impact of different architectural decisions of our $\encodec$ model and we present our results with objective metrics and real-time factor in Table~\ref{tab:arch}. For all models, we consider the streamable setup and our reference $\encodec$ base model has the number of channels $C$ set to 32 and a single residual unit. The real-time factor reported here is defined as the ratio between the duration of the input audio and the processing time needed for encoding (respectively decoding) it. We profiled streamable multi-target 24 kHz models during inference at 6 kbps on a single thread of a MacBook Pro 2019 CPU. 
First, we validate that the selected number of channels provide good trade offs in terms of perceived quality and inference speed. We observe that increasing the capacity of the model only marginally affects the scores on objective metrics while it has a high impact on the real-time factor. 
The results also demonstrate that presence of LSTM improves the SI-SNR and the final reconstruction quality, at the detriment of the real-time factor. We also experiment with increasing the number of residual units instead of relying on a LSTM in our architecture. To do so, we use 3 Residual Units and we double the dilation used in the first convolutional layer of the residual unit for each subsequent unit. We observe that using Residual Units has more impact on the real-time factor and we note a small degradation of the SI-SNR compared to the LSTM-based version.

\begin{table}[t!]
  \caption{Model architecture Analysis. We explore variations of our architecture including the impact of the number of Residual Units (ResUnits), the sequence modeling with LSTM, the number of channels (C) on objective metrics and real-time factor (RTF greater than 1 is faster than real time).
  \label{tab:arch}}
  \centering
  \begin{tabular}{lrrcc}
    \toprule
    Model           & RTF Enc & RTF Dec & SI-SNR & ViSQOL \\
    \midrule
    $\encodec$ base              & 9.8 & 10.4 & 6.67 & 4.35\\
    \midrule
    Channels = 16                & 26.0 & 25.7 & 6.40 & 4.32\\
    Channels = 64                &  1.3 &  3.1 & 6.70 & 4.38\\
    norm = None                  & 10.1 & 10.4 & 6.45 & 4.29\\
    LSTM = 0                     & 15.0 & 14.6 & 6.40 & 4.35\\
    Residual Layer = 3, LSTM = 0 & 6.0 & 7.3 & 6.32 & 4.35\\
    \bottomrule
  \end{tabular}
\end{table}

{\noindent \bf The effect of the balancer.}
Lastly, we present results evaluating the impact of the balancer. In which, we train the $\encodec$ model considering various levels of balancing the loss. For this set of experiments we use the DiffQ quantizer other RVQ. All models were trained on Jamando music dataset (see Table~\ref{tab:balancer}). Results suggest the balancer significantly stabilizes the training process. This is especially useful while considering the different terms in the objective together where each term operates at a different scale. Following the balancer approach significantly reduce the effort needed for tuning the objective coefficients (i.e., $\lambda_t$, $\lambda_f$, $\lambda_g$, $\lambda_{feat}$). We demonstrate that the use of the balancer shows no degradation compared to an identified combination of coefficients.

\begin{table}[t!]
    \centering
    \caption{ViSQOL and SI-SNR results for $\encodec$ using DiffQ considering various coefficients to balance the overall objective. All models were trained using Jamando music dataset.}
    \label{tab:balancer}
    \begin{tabular}{lllll|rr}
        \toprule
        $\lambda_t$ & $\lambda_f$ & $\lambda_{g}$ & $\lambda_{feat}$ & Balancer & SI-SNR & ViSQOL \\
        \midrule
        1 & 1 & 1 & 1 & \ding{51} & \bf 10.32 & \bf 4.16\\
        1 & 1 & 1 & 1 & \ding{55} & 6.16  & 3.89\\
        \midrule
        1 & 1 & 2 & 1 & \ding{51} & \bf 10.08 & \bf 4.12\\
        1 & 1 & 2 & 1 & \ding{55} & 5.01  & 3.77\\
        \midrule
        1 & 2 & 2 & 1 & \ding{51} & \bf 10.06 & \bf 4.17\\
        1 & 2 & 2 & 1 & \ding{55} & 3.84  & 3.67\\
        \midrule
        1 & 2 & 4 & 1 & \ding{51} & \bf 9.93   & \bf 4.17\\
        1 & 2 & 4 & 1 & \ding{55} & 1.72  & 3.52\\
        \midrule
        1 & 2 & 100 & 1 & \ding{51} & \bf 8.41 & \bf 4.05\\
        1 & 2 & 100 & 1 & \ding{55} & -35.83& 2.82\\
        \midrule
        2 & 1 & 1 & 4 & \ding{51} & \bf 10.53 & \bf 4.06\\
        2 & 1 & 1 & 4 & \ding{55} & 7.66 & 4.03\\
        \midrule
        2 & 2 & 2 & 4 & \ding{51} & \bf 10.19 & \bf 4.13\\
        2 & 2 & 2 & 4 & \ding{55} & 7.16 & 3.98\\
        \midrule
        2 & 2 & 10 & 4 & \ding{51} & \bf 9.82 & \bf 4.11\\
        2 & 2 & 10 & 4 & \ding{55} & 3.98 & 3.66\\
        \midrule
        2 & 2 & 100 & 4 & \ding{51} &  \bf 8.52 & \bf 4.03\\
        2 & 2 & 100 & 4 & \ding{55} & -34.31 & 2.91\\
        \midrule
        10 & 1 & 1 & 1 & \ding{51} & \bf 10.53 & 3.65\\
        10 & 1 & 1 & 1 & \ding{55} & 8.16 & \bf 4.00\\
        \midrule
        10 & 1 & 4 & 1 & \ding{51} & \bf 10.72 & 3.62\\
        10 & 1 & 4 & 1 & \ding{55} &  5.99 & \bf 3.74\\
        \midrule
        10 & 1 & 4 & 2 & \ding{51} & \bf 10.71 & 3.73\\
        10 & 1 & 4 & 2 & \ding{55} & 7.03 & \bf 3.88\\
        \midrule
        10 & 2 & 100 & 2 & \ding{51} & \bf 9.23 & \bf 4.07\\
        10 & 2 & 100 & 2 & \ding{55} & -33.78 & 2.85\\
        \midrule
        10 & 2 & 100 & 4 & \ding{51} & \bf 9.22 & \bf 4.09\\
        10 & 2 & 100 & 4 & \ding{55} & -16.39 & 2.95\\
        \bottomrule
    \end{tabular}
\end{table}
\subsection{Societal impact}
\label{societal_impact}

The majority of the internet traffic is represented by audio and video streams (82\% in 2021 according to~\citet{internet_traffic_cisco}). This share of content is boosted by user-generated content, phone and video calls and by the development of HD music streaming and video streaming services. Compression methods are used to reduce the storage requirements and network bandwidth used to serve this content. Furthermore, the growing adoption of wearable devices contributes to making efficient compression an increasingly important problem. 

Different audio codecs, including Opus and EVS, have been developed and widely adopted over the past years. Those codecs support audio coding at low latency with high audio quality at low to medium bitrates (in the range of 12 to 24 kbps) but the audio quality deteriorates at very low bitrates (eg. 3 kbps) on non-speech audio. Addressing very low bitrate compression with high fidelity remains an essential challenge to solve as very low bitrate codecs enable communication and improve experiences such as videoconferencing or streaming content even with a poor internet connection and therefore allows the internet services to become more inclusive. While further work needs to be done, we hope that sharing the result of this work to the broader community can further contribute to this direction.

%% file: 00_main.bbl
\begin{thebibliography}{76}
\providecommand{\natexlab}[1]{#1}
\providecommand{\url}[1]{\texttt{#1}}
\expandafter\ifx\csname urlstyle\endcsname\relax
  \providecommand{\doi}[1]{doi: #1}\else
  \providecommand{\doi}{doi: \begingroup \urlstyle{rm}\Url}\fi

\bibitem[Andreev et~al.(2022)Andreev, Alanov, Ivanov, and Vetrov]{hifi++}
Pavel Andreev, Aibek Alanov, Oleg Ivanov, and Dmitry Vetrov.
\newblock Hifi++: a unified framework for neural vocoding, bandwidth extension
  and speech enhancement.
\newblock \emph{arXiv preprint arXiv:2203.13086}, 2022.

\bibitem[Ardila et~al.(2019)Ardila, Branson, Davis, Henretty, Kohler, Meyer,
  Morais, Saunders, Tyers, and Weber]{ardila2019common}
Rosana Ardila, Megan Branson, Kelly Davis, Michael Henretty, Michael Kohler,
  Josh Meyer, Reuben Morais, Lindsay Saunders, Francis~M Tyers, and Gregor
  Weber.
\newblock Common voice: A massively-multilingual speech corpus.
\newblock \emph{arXiv preprint arXiv:1912.06670}, 2019.

\bibitem[Atal \& Hanauer(1971)Atal and Hanauer]{atal1971speech}
Bishnu~S Atal and Suzanne~L Hanauer.
\newblock Speech analysis and synthesis by linear prediction of the speech
  wave.
\newblock \emph{The journal of the acoustical society of America}, 50\penalty0
  (2B):\penalty0 637--655, 1971.

\bibitem[Ba et~al.(2016)Ba, Kiros, and Hinton]{ba2016layer}
Jimmy~Lei Ba, Jamie~Ryan Kiros, and Geoffrey~E Hinton.
\newblock Layer normalization.
\newblock \emph{arXiv preprint arXiv:1607.06450}, 2016.

\bibitem[Ball{\'e} et~al.(2017)Ball{\'e}, Laparra, and
  Simoncelli]{balle2017end}
Johannes Ball{\'e}, Valero Laparra, and Eero~P Simoncelli.
\newblock End-to-end optimized image compression.
\newblock In \emph{ICLR}, 2017.

\bibitem[Ball{\'e} et~al.(2018)Ball{\'e}, Johnston, and
  Minnen]{balle2018integer}
Johannes Ball{\'e}, Nick Johnston, and David Minnen.
\newblock Integer networks for data compression with latent-variable models.
\newblock In \emph{International Conference on Learning Representations}, 2018.

\bibitem[Bengio et~al.(2013)Bengio, L{\'e}onard, and
  Courville]{bengio2013estimating}
Yoshua Bengio, Nicholas L{\'e}onard, and Aaron Courville.
\newblock Estimating or propagating gradients through stochastic neurons for
  conditional computation.
\newblock \emph{arXiv preprint arXiv:1308.3432}, 2013.

\bibitem[Bhagat et~al.(2012)Bhagat, Bhatt, and Kosta]{bhagat2012adaptive}
Dipesh Bhagat, Ninad Bhatt, and Yogeshwar Kosta.
\newblock Adaptive multi-rate wideband speech codec based on celp algorithm:
  architectural study, implementation \& performance analysis.
\newblock In \emph{2012 International Conference on Communication Systems and
  Network Technologies}, pp.\  547--551. IEEE, 2012.

\bibitem[Bogdanov et~al.(2019)Bogdanov, Won, Tovstogan, Porter, and
  Serra]{bogdanov2019mtg}
Dmitry Bogdanov, Minz Won, Philip Tovstogan, Alastair Porter, and Xavier Serra.
\newblock The mtg-jamendo dataset for automatic music tagging.
\newblock In \emph{Machine Learning for Music Discovery Workshop, International
  Conference on Machine Learning (ICML 2019)}, Long Beach, CA, United States,
  2019.
\newblock URL \url{http://hdl.handle.net/10230/42015}.

\bibitem[Chazan et~al.(2021)Chazan, Wolf, Nachmani, and Adi]{chazan2021single}
Shlomo~E Chazan, Lior Wolf, Eliya Nachmani, and Yossi Adi.
\newblock Single channel voice separation for unknown number of speakers under
  reverberant and noisy settings.
\newblock In \emph{ICASSP 2021-2021 IEEE International Conference on Acoustics,
  Speech and Signal Processing (ICASSP)}, pp.\  3730--3734. IEEE, 2021.

\bibitem[Chinen et~al.(2020)Chinen, Lim, Skoglund, Gureev, O'Gorman, and
  Hines]{chinen2020visqol}
Michael Chinen, Felicia~SC Lim, Jan Skoglund, Nikita Gureev, Feargus O'Gorman,
  and Andrew Hines.
\newblock Visqol v3: An open source production ready objective speech and audio
  metric.
\newblock In \emph{2020 twelfth international conference on quality of
  multimedia experience (QoMEX)}, pp.\  1--6. IEEE, 2020.

\bibitem[Cisco(2021)]{internet_traffic_cisco}
Cisco.
\newblock Global - 2021 forecast highlights - cisco.
\newblock
  \url{https://www.cisco.com/c/dam/m/en_us/solutions/service-provider/vni-forecast-highlights/pdf/Global_2021_Forecast_Highlights.pdf},
  2021.

\bibitem[Clevert et~al.(2015)Clevert, Unterthiner, and
  Hochreiter]{clevert2015fast}
Djork-Arn{\'e} Clevert, Thomas Unterthiner, and Sepp Hochreiter.
\newblock Fast and accurate deep network learning by exponential linear units
  (elus).
\newblock \emph{arXiv preprint arXiv:1511.07289}, 2015.

\bibitem[D{\'e}fossez et~al.(2019)D{\'e}fossez, Usunier, Bottou, and
  Bach]{defossez2019music}
Alexandre D{\'e}fossez, Nicolas Usunier, L{\'e}on Bottou, and Francis Bach.
\newblock Music source separation in the waveform domain.
\newblock \emph{arXiv preprint arXiv:1911.13254}, 2019.

\bibitem[Defossez et~al.(2020)Defossez, Synnaeve, and Adi]{defossez2020real}
Alexandre Defossez, Gabriel Synnaeve, and Yossi Adi.
\newblock Real time speech enhancement in the waveform domain.
\newblock \emph{arXiv preprint arXiv:2006.12847}, 2020.

\bibitem[D{\'e}fossez et~al.(2021)D{\'e}fossez, Adi, and
  Synnaeve]{defossez2021differentiable}
Alexandre D{\'e}fossez, Yossi Adi, and Gabriel Synnaeve.
\newblock Differentiable model compression via pseudo quantization noise.
\newblock \emph{arXiv preprint arXiv:2104.09987}, 2021.

\bibitem[Dhariwal et~al.(2020)Dhariwal, Jun, Payne, Kim, Radford, and
  Sutskever]{dhariwal2020jukebox}
Prafulla Dhariwal, Heewoo Jun, Christine Payne, Jong~Wook Kim, Alec Radford,
  and Ilya Sutskever.
\newblock Jukebox: A generative model for music.
\newblock \emph{arXiv preprint arXiv:2005.00341}, 2020.

\bibitem[Dieleman et~al.(2018)Dieleman, van~den Oord, and
  Simonyan]{dieleman2018challenge}
Sander Dieleman, Aaron van~den Oord, and Karen Simonyan.
\newblock The challenge of realistic music generation: modelling raw audio at
  scale.
\newblock \emph{Advances in Neural Information Processing Systems}, 31, 2018.

\bibitem[Dietz et~al.(2015)Dietz, Multrus, Eksler, Malenovsky, Norvell,
  Pobloth, Miao, Wang, Laaksonen, Vasilache, et~al.]{dietz2015overview}
Martin Dietz, Markus Multrus, Vaclav Eksler, Vladimir Malenovsky, Erik Norvell,
  Harald Pobloth, Lei Miao, Zhe Wang, Lasse Laaksonen, Adriana Vasilache,
  et~al.
\newblock Overview of the evs codec architecture.
\newblock In \emph{2015 IEEE International Conference on Acoustics, Speech and
  Signal Processing (ICASSP)}, pp.\  5698--5702. IEEE, 2015.

\bibitem[Dubey et~al.(2022)Dubey, Gopal, Cutler, Matusevych, Braun, Eskimez,
  Thakker, Yoshioka, Gamper, and Aichner]{dns}
Harishchandra Dubey, Vishak Gopal, Ross Cutler, Sergiy Matusevych, Sebastian
  Braun, Emre~Sefik Eskimez, Manthan Thakker, Takuya Yoshioka, Hannes Gamper,
  and Robert Aichner.
\newblock Icassp 2022 deep noise suppression challenge.
\newblock In \emph{ICASSP}, 2022.

\bibitem[Fonseca et~al.(2021)Fonseca, Favory, Pons, Font, and
  Serra]{fonseca2021fsd50k}
Eduardo Fonseca, Xavier Favory, Jordi Pons, Frederic Font, and Xavier Serra.
\newblock Fsd50k: an open dataset of human-labeled sound events.
\newblock \emph{IEEE/ACM Transactions on Audio, Speech, and Language
  Processing}, 30:\penalty0 829--852, 2021.

\bibitem[G{\^a}rbacea et~al.(2019)G{\^a}rbacea, van~den Oord, Li, Lim, Luebs,
  Vinyals, and Walters]{garbacea2019low}
Cristina G{\^a}rbacea, A{\"a}ron van~den Oord, Yazhe Li, Felicia~SC Lim,
  Alejandro Luebs, Oriol Vinyals, and Thomas~C Walters.
\newblock Low bit-rate speech coding with vq-vae and a wavenet decoder.
\newblock In \emph{ICASSP 2019-2019 IEEE International Conference on Acoustics,
  Speech and Signal Processing (ICASSP)}, pp.\  735--739. IEEE, 2019.

\bibitem[Gemmeke et~al.(2017)Gemmeke, Ellis, Freedman, Jansen, Lawrence, Moore,
  Plakal, and Ritter]{gemmeke2017audio}
Jort~F Gemmeke, Daniel~PW Ellis, Dylan Freedman, Aren Jansen, Wade Lawrence,
  R~Channing Moore, Manoj Plakal, and Marvin Ritter.
\newblock Audio set: An ontology and human-labeled dataset for audio events.
\newblock In \emph{2017 IEEE international conference on acoustics, speech and
  signal processing (ICASSP)}, pp.\  776--780. IEEE, 2017.

\bibitem[Goel et~al.(2022)Goel, Gu, Donahue, and R{\'e}]{goel2022s}
Karan Goel, Albert Gu, Chris Donahue, and Christopher R{\'e}.
\newblock It's raw! audio generation with state-space models.
\newblock \emph{arXiv preprint arXiv:2202.09729}, 2022.

\bibitem[Gray(1984)]{gray1984vector}
Robert Gray.
\newblock Vector quantization.
\newblock \emph{IEEE Assp Magazine}, 1\penalty0 (2):\penalty0 4--29, 1984.

\bibitem[Griffin \& Lim(1985)Griffin and Lim]{griffin1985new}
D~Griffin and Jae Lim.
\newblock A new model-based speech analysis/synthesis system.
\newblock In \emph{ICASSP}, 1985.

\bibitem[Gritsenko et~al.(2020)Gritsenko, Salimans, van~den Berg, Snoek, and
  Kalchbrenner]{gritsenko2020spectral}
Alexey Gritsenko, Tim Salimans, Rianne van~den Berg, Jasper Snoek, and Nal
  Kalchbrenner.
\newblock A spectral energy distance for parallel speech synthesis.
\newblock \emph{Advances in Neural Information Processing Systems},
  33:\penalty0 13062--13072, 2020.

\bibitem[Hines et~al.(2012)Hines, Skoglund, Kokaram, and
  Harte]{hines2012visqol}
Andrew Hines, Jan Skoglund, Anil Kokaram, and Naomi Harte.
\newblock Visqol: The virtual speech quality objective listener.
\newblock In \emph{IWAENC 2012; International Workshop on Acoustic Signal
  Enhancement}, pp.\  1--4. VDE, 2012.

\bibitem[Hsu et~al.(2021)Hsu, Bolte, Tsai, Lakhotia, Salakhutdinov, and
  Mohamed]{hsu2021hubert}
Wei-Ning Hsu, Benjamin Bolte, Yao-Hung~Hubert Tsai, Kushal Lakhotia, Ruslan
  Salakhutdinov, and Abdelrahman Mohamed.
\newblock Hubert: Self-supervised speech representation learning by masked
  prediction of hidden units.
\newblock \emph{IEEE/ACM Transactions on Audio, Speech, and Language
  Processing}, 29:\penalty0 3451--3460, 2021.

\bibitem[Ioffe \& Szegedy(2015)Ioffe and Szegedy]{batchnorm}
Sergey Ioffe and Christian Szegedy.
\newblock Batch normalization: Accelerating deep network training by reducing
  internal covariate shift.
\newblock Technical Report 1502.03167, arXiv, 2015.

\bibitem[Jang et~al.(2017)Jang, Gu, and Poole]{jang2016categorical}
Eric Jang, Shixiang Gu, and Ben Poole.
\newblock Categorical reparameterization with gumbel-softmax.
\newblock In \emph{ICLR}, 2017.

\bibitem[Jayashankar et~al.(2022)Jayashankar, Koehler, Kalgaonkar, Xiu, Wu,
  Lin, Agrawal, and He]{jayashankar2022architecture}
Tejas Jayashankar, Thilo Koehler, Kaustubh Kalgaonkar, Zhiping Xiu, Jilong Wu,
  Ju~Lin, Prabhav Agrawal, and Qing He.
\newblock Architecture for variable bitrate neural speech codec with
  configurable computation complexity.
\newblock In \emph{ICASSP 2022-2022 IEEE International Conference on Acoustics,
  Speech and Signal Processing (ICASSP)}, pp.\  861--865. IEEE, 2022.

\bibitem[Jiang et~al.(2022)Jiang, Peng, Zheng, Xue, Zhang, and
  Lu]{jiang2022end}
Xue Jiang, Xiulian Peng, Chengyu Zheng, Huaying Xue, Yuan Zhang, and Yan Lu.
\newblock End-to-end neural speech coding for real-time communications.
\newblock In \emph{ICASSP 2022-2022 IEEE International Conference on Acoustics,
  Speech and Signal Processing (ICASSP)}, pp.\  866--870. IEEE, 2022.

\bibitem[Juang \& Gray(1982)Juang and Gray]{juang1982multiple}
Biing-Hwang Juang and A~Gray.
\newblock Multiple stage vector quantization for speech coding.
\newblock In \emph{ICASSP'82. IEEE International Conference on Acoustics,
  Speech, and Signal Processing}, volume~7, pp.\  597--600. IEEE, 1982.

\bibitem[Kalchbrenner et~al.(2018)]{wavernn}
Nal Kalchbrenner et~al.
\newblock {Efficient} {Neural} {Audio} {Synthesis}.
\newblock In \emph{ICML}, 2018.

\bibitem[Kharitonov et~al.(2021)Kharitonov, Lee, Polyak, Adi, Copet, Lakhotia,
  Nguyen, Rivi{\`e}re, Mohamed, Dupoux, et~al.]{kharitonov2021text}
Eugene Kharitonov, Ann Lee, Adam Polyak, Yossi Adi, Jade Copet, Kushal
  Lakhotia, Tu-Anh Nguyen, Morgane Rivi{\`e}re, Abdelrahman Mohamed, Emmanuel
  Dupoux, et~al.
\newblock Text-free prosody-aware generative spoken language modeling.
\newblock \emph{arXiv preprint arXiv:2109.03264}, 2021.

\bibitem[Kleijn et~al.(2018)Kleijn, Lim, Luebs, Skoglund, Stimberg, Wang, and
  Walters]{kleijn2018wavenet}
W~Bastiaan Kleijn, Felicia~SC Lim, Alejandro Luebs, Jan Skoglund, Florian
  Stimberg, Quan Wang, and Thomas~C Walters.
\newblock Wavenet based low rate speech coding.
\newblock In \emph{2018 IEEE international conference on acoustics, speech and
  signal processing (ICASSP)}, pp.\  676--680. IEEE, 2018.

\bibitem[Kleijn et~al.(2021)Kleijn, Storus, Chinen, Denton, Lim, Luebs,
  Skoglund, and Yeh]{kleijn2021generative}
W~Bastiaan Kleijn, Andrew Storus, Michael Chinen, Tom Denton, Felicia~SC Lim,
  Alejandro Luebs, Jan Skoglund, and Hengchin Yeh.
\newblock Generative speech coding with predictive variance regularization.
\newblock In \emph{ICASSP 2021-2021 IEEE International Conference on Acoustics,
  Speech and Signal Processing (ICASSP)}, pp.\  6478--6482. IEEE, 2021.

\bibitem[Kong et~al.(2020)Kong, Kim, and Bae]{hifigan}
Jungil Kong, Jaehyeon Kim, and Jaekyoung Bae.
\newblock Hifi-gan: Generative adversarial networks for efficient and high
  fidelity speech synthesis.
\newblock \emph{Advances in Neural Information Processing Systems},
  33:\penalty0 17022--17033, 2020.

\bibitem[Kreuk et~al.(2021)Kreuk, Polyak, Copet, Kharitonov, Nguyen,
  Rivi{\`e}re, Hsu, Mohamed, Dupoux, and Adi]{kreuk2021textless}
Felix Kreuk, Adam Polyak, Jade Copet, Eugene Kharitonov, Tu-Anh Nguyen, Morgane
  Rivi{\`e}re, Wei-Ning Hsu, Abdelrahman Mohamed, Emmanuel Dupoux, and Yossi
  Adi.
\newblock Textless speech emotion conversion using decomposed and discrete
  representations.
\newblock \emph{arXiv preprint arXiv:2111.07402}, 2021.

\bibitem[Kumar et~al.(2019)Kumar, Kumar, de~Boissiere, Gestin, Teoh, Sotelo,
  de~Br{\'e}bisson, Bengio, and Courville]{melgan}
Kundan Kumar, Rithesh Kumar, Thibault de~Boissiere, Lucas Gestin, Wei~Zhen
  Teoh, Jose Sotelo, Alexandre de~Br{\'e}bisson, Yoshua Bengio, and Aaron~C
  Courville.
\newblock Melgan: Generative adversarial networks for conditional waveform
  synthesis.
\newblock \emph{Advances in neural information processing systems}, 32, 2019.

\bibitem[Lakhotia et~al.(2021)Lakhotia, Kharitonov, Hsu, Adi, Polyak, Bolte,
  Nguyen, Copet, Baevski, Mohamed, et~al.]{lakhotia2021generative}
Kushal Lakhotia, Eugene Kharitonov, Wei-Ning Hsu, Yossi Adi, Adam Polyak,
  Benjamin Bolte, Tu-Anh Nguyen, Jade Copet, Alexei Baevski, Abdelrahman
  Mohamed, et~al.
\newblock On generative spoken language modeling from raw audio.
\newblock \emph{Transactions of the Association for Computational Linguistics},
  9:\penalty0 1336--1354, 2021.

\bibitem[Lee et~al.(2021{\natexlab{a}})Lee, Chen, Wang, Gu, Ma, Polyak, Adi,
  He, Tang, Pino, et~al.]{lee2021direct}
Ann Lee, Peng-Jen Chen, Changhan Wang, Jiatao Gu, Xutai Ma, Adam Polyak, Yossi
  Adi, Qing He, Yun Tang, Juan Pino, et~al.
\newblock Direct speech-to-speech translation with discrete units.
\newblock \emph{arXiv preprint arXiv:2107.05604}, 2021{\natexlab{a}}.

\bibitem[Lee et~al.(2021{\natexlab{b}})Lee, Gong, Duquenne, Schwenk, Chen,
  Wang, Popuri, Pino, Gu, and Hsu]{lee2021textless}
Ann Lee, Hongyu Gong, Paul-Ambroise Duquenne, Holger Schwenk, Peng-Jen Chen,
  Changhan Wang, Sravya Popuri, Juan Pino, Jiatao Gu, and Wei-Ning Hsu.
\newblock Textless speech-to-speech translation on real data.
\newblock \emph{arXiv preprint arXiv:2112.08352}, 2021{\natexlab{b}}.

\bibitem[Li et~al.()Li, Mao, and McAuley]{livariable}
Shuyang Li, Huanru~Henry Mao, and Julian McAuley.
\newblock Variable bitrate discrete neural representations via causal
  self-attention.
\newblock In \emph{2nd Pre-registration workshop (NeurIPS 2021), Remote.}

\bibitem[Li et~al.(2021)Li, Tagliasacchi, Rybakov, Ungureanu, and
  Roblek]{li2021real}
Yunpeng Li, Marco Tagliasacchi, Oleg Rybakov, Victor Ungureanu, and Dominik
  Roblek.
\newblock Real-time speech frequency bandwidth extension.
\newblock In \emph{ICASSP 2021-2021 IEEE International Conference on Acoustics,
  Speech and Signal Processing (ICASSP)}, pp.\  691--695. IEEE, 2021.

\bibitem[Lim et~al.(2020)Lim, Kleijn, Chinen, and Skoglund]{lim2020robust}
Felicia~SC Lim, W~Bastiaan Kleijn, Michael Chinen, and Jan Skoglund.
\newblock Robust low rate speech coding based on cloned networks and wavenet.
\newblock In \emph{ICASSP 2020-2020 IEEE International Conference on Acoustics,
  Speech and Signal Processing (ICASSP)}, pp.\  6769--6773. IEEE, 2020.

\bibitem[Lin et~al.(2022)Lin, Kalgaonkar, He, and Lei]{lin2022speech}
Ju~Lin, Kaustubh Kalgaonkar, Qing He, and Xin Lei.
\newblock Speech enhancement for low bit rate speech codec.
\newblock In \emph{ICASSP 2022-2022 IEEE International Conference on Acoustics,
  Speech and Signal Processing (ICASSP)}, pp.\  7777--7781. IEEE, 2022.

\bibitem[Luo \& Mesgarani(2019)Luo and Mesgarani]{luo2019conv}
Yi~Luo and Nima Mesgarani.
\newblock Conv-tasnet: Surpassing ideal time--frequency magnitude masking for
  speech separation.
\newblock \emph{IEEE/ACM transactions on audio, speech, and language
  processing}, 27\penalty0 (8):\penalty0 1256--1266, 2019.

\bibitem[McCree et~al.(1996)McCree, Truong, George, Barnwell, and
  Viswanathan]{mccree19962}
Alan McCree, Kwan Truong, E~Bryan George, Thomas~P Barnwell, and Vishu
  Viswanathan.
\newblock A 2.4 kbit/s melp coder candidate for the new us federal standard.
\newblock In \emph{ICASSP}, 1996.

\bibitem[Morishima et~al.(1990)Morishima, Harashima, and
  Katayama]{morishima1990speech}
Shigeo Morishima, H~Harashima, and Y~Katayama.
\newblock Speech coding based on a multi-layer neural network.
\newblock In \emph{IEEE International Conference on Communications, Including
  Supercomm Technical Sessions}, pp.\  429--433. IEEE, 1990.

\bibitem[Nachmani et~al.(2020)Nachmani, Adi, and Wolf]{nachmani2020voice}
Eliya Nachmani, Yossi Adi, and Lior Wolf.
\newblock Voice separation with an unknown number of multiple speakers.
\newblock In \emph{International Conference on Machine Learning}, pp.\
  7164--7175. PMLR, 2020.

\bibitem[Nguyen et~al.(2022)Nguyen, Kharitonov, Copet, Adi, Hsu, Elkahky,
  Tomasello, Algayres, Sagot, Mohamed, et~al.]{nguyen2022generative}
Tu~Anh Nguyen, Eugene Kharitonov, Jade Copet, Yossi Adi, Wei-Ning Hsu, Ali
  Elkahky, Paden Tomasello, Robin Algayres, Benoit Sagot, Abdelrahman Mohamed,
  et~al.
\newblock Generative spoken dialogue language modeling.
\newblock \emph{arXiv preprint arXiv:2203.16502}, 2022.

\bibitem[Omran et~al.(2022)Omran, Zeghidour, Borsos, Quitry, Slaney, and
  Tagliasacchi]{omran2022disentangling}
Ahmed Omran, Neil Zeghidour, Zal{\'a}n Borsos, F{\'e}lix de~Chaumont Quitry,
  Malcolm Slaney, and Marco Tagliasacchi.
\newblock Disentangling speech from surroundings in a neural audio codec.
\newblock \emph{arXiv preprint arXiv:2203.15578}, 2022.

\bibitem[Oord et~al.(2016)Oord, Dieleman, Zen, Simonyan, Vinyals, Graves,
  Kalchbrenner, Senior, and Kavukcuoglu]{oord2016wavenet}
Aaron van~den Oord, Sander Dieleman, Heiga Zen, Karen Simonyan, Oriol Vinyals,
  Alex Graves, Nal Kalchbrenner, Andrew Senior, and Koray Kavukcuoglu.
\newblock Wavenet: A generative model for raw audio.
\newblock \emph{arXiv preprint arXiv:1609.03499}, 2016.

\bibitem[Pasco(1976)]{pasco1976source}
Richard~Clark Pasco.
\newblock \emph{Source coding algorithms for fast data compression}.
\newblock PhD thesis, Stanford University CA, 1976.

\bibitem[Polyak et~al.(2021)Polyak, Adi, Copet, Kharitonov, Lakhotia, Hsu,
  Mohamed, and Dupoux]{polyak2021speech}
Adam Polyak, Yossi Adi, Jade Copet, Eugene Kharitonov, Kushal Lakhotia,
  Wei-Ning Hsu, Abdelrahman Mohamed, and Emmanuel Dupoux.
\newblock Speech resynthesis from discrete disentangled self-supervised
  representations.
\newblock \emph{arXiv preprint arXiv:2104.00355}, 2021.

\bibitem[Popuri et~al.(2022)Popuri, Chen, Wang, Pino, Adi, Gu, Hsu, and
  Lee]{popuri2022enhanced}
Sravya Popuri, Peng-Jen Chen, Changhan Wang, Juan Pino, Yossi Adi, Jiatao Gu,
  Wei-Ning Hsu, and Ann Lee.
\newblock Enhanced direct speech-to-speech translation using self-supervised
  pre-training and data augmentation.
\newblock \emph{arXiv preprint arXiv:2204.02967}, 2022.

\bibitem[Razavi et~al.(2019)Razavi, Van~den Oord, and
  Vinyals]{razavi2019generating}
Ali Razavi, Aaron Van~den Oord, and Oriol Vinyals.
\newblock Generating diverse high-fidelity images with vq-vae-2.
\newblock \emph{Advances in neural information processing systems}, 32, 2019.

\bibitem[Rippel et~al.(2019)Rippel, Nair, Lew, Branson, Anderson, and
  Bourdev]{rippel2019learned}
Oren Rippel, Sanjay Nair, Carissa Lew, Steve Branson, Alexander~G Anderson, and
  Lubomir Bourdev.
\newblock Learned video compression.
\newblock In \emph{Proceedings of the IEEE/CVF International Conference on
  Computer Vision}, pp.\  3454--3463, 2019.

\bibitem[Rissanen \& Langdon(1981)Rissanen and Langdon]{rissanen1981universal}
Jorma Rissanen and Glen Langdon.
\newblock Universal modeling and coding.
\newblock \emph{IEEE Transactions on Information Theory}, 27\penalty0
  (1):\penalty0 12--23, 1981.

\bibitem[Salimans \& Kingma(2016)Salimans and Kingma]{salimans2016weight}
Tim Salimans and Durk~P Kingma.
\newblock Weight normalization: A simple reparameterization to accelerate
  training of deep neural networks.
\newblock \emph{Advances in neural information processing systems}, 29, 2016.

\bibitem[Series(2014)]{series2014method}
B~Series.
\newblock Method for the subjective assessment of intermediate quality level of
  audio systems.
\newblock \emph{International Telecommunication Union Radiocommunication
  Assembly}, 2014.

\bibitem[Skoglund \& Valin(2019)Skoglund and Valin]{skoglund2019improving}
Jan Skoglund and Jean-Marc Valin.
\newblock Improving opus low bit rate quality with neural speech synthesis.
\newblock \emph{arXiv preprint arXiv:1905.04628}, 2019.

\bibitem[Tagliasacchi et~al.(2020)Tagliasacchi, Li, Misiunas, and
  Roblek]{tagliasacchi2020seanet}
Marco Tagliasacchi, Yunpeng Li, Karolis Misiunas, and Dominik Roblek.
\newblock Seanet: A multi-modal speech enhancement network.
\newblock \emph{arXiv preprint arXiv:2009.02095}, 2020.

\bibitem[Valin \& Skoglund(2019{\natexlab{a}})Valin and
  Skoglund]{valin2019lpcnet}
Jean-Marc Valin and Jan Skoglund.
\newblock Lpcnet: Improving neural speech synthesis through linear prediction.
\newblock In \emph{ICASSP 2019-2019 IEEE International Conference on Acoustics,
  Speech and Signal Processing (ICASSP)}, pp.\  5891--5895. IEEE,
  2019{\natexlab{a}}.

\bibitem[Valin \& Skoglund(2019{\natexlab{b}})Valin and
  Skoglund]{valin2019real}
Jean-Marc Valin and Jan Skoglund.
\newblock A real-time wideband neural vocoder at 1.6 kb/s using lpcnet.
\newblock \emph{arXiv preprint arXiv:1903.12087}, 2019{\natexlab{b}}.

\bibitem[Valin et~al.(2012)Valin, Vos, and Terriberry]{valin2012definition}
Jean-Marc Valin, Koen Vos, and Timothy Terriberry.
\newblock Definition of the opus audio codec.
\newblock \emph{IETF, September}, 2, 2012.

\bibitem[Van Den~Oord et~al.(2017)Van Den~Oord, Vinyals, et~al.]{van2017neural}
Aaron Van Den~Oord, Oriol Vinyals, et~al.
\newblock Neural discrete representation learning.
\newblock \emph{Advances in neural information processing systems}, 30, 2017.

\bibitem[Vasuki \& Vanathi(2006)Vasuki and Vanathi]{vasuki2006review}
A~Vasuki and PT~Vanathi.
\newblock A review of vector quantization techniques.
\newblock \emph{IEEE Potentials}, 25\penalty0 (4):\penalty0 39--47, 2006.

\bibitem[Vaswani et~al.(2017)Vaswani, Shazeer, Parmar, Uszkoreit, Jones, Gomez,
  Kaiser, and Polosukhin]{vaswani2017attention}
Ashish Vaswani, Noam Shazeer, Niki Parmar, Jakob Uszkoreit, Llion Jones,
  Aidan~N Gomez, Lukasz Kaiser, and Illia Polosukhin.
\newblock Attention is all you need.
\newblock In \emph{Proc. of Neural Information Processing Systems}, 2017.

\bibitem[Widrow et~al.(1996)Widrow, Kollar, and Liu]{widrow1996statistical}
Bernard Widrow, Istvan Kollar, and Ming-Chang Liu.
\newblock Statistical theory of quantization.
\newblock \emph{IEEE Transactions on instrumentation and measurement},
  45\penalty0 (2):\penalty0 353--361, 1996.

\bibitem[Yamamoto et~al.(2020{\natexlab{a}})Yamamoto, Song, and
  Kim]{parallelwavegan}
Ryuichi Yamamoto, Eunwoo Song, and Jae-Min Kim.
\newblock Parallel wavegan: A fast waveform generation model based on
  generative adversarial networks with multi-resolution spectrogram.
\newblock In \emph{ICASSP 2020-2020 IEEE International Conference on Acoustics,
  Speech and Signal Processing (ICASSP)}, pp.\  6199--6203. IEEE,
  2020{\natexlab{a}}.

\bibitem[Yamamoto et~al.(2020{\natexlab{b}})Yamamoto, Song, and
  Kim]{yamamoto2020parallel}
Ryuichi Yamamoto, Eunwoo Song, and Jae-Min Kim.
\newblock Parallel wavegan: A fast waveform generation model based on
  generative adversarial networks with multi-resolution spectrogram.
\newblock In \emph{ICASSP 2020-2020 IEEE International Conference on Acoustics,
  Speech and Signal Processing (ICASSP)}, pp.\  6199--6203. IEEE,
  2020{\natexlab{b}}.

\bibitem[You et~al.(2021)You, Kim, Nam, Hwang, and Chae]{you2021gan}
Jaeseong You, Dalhyun Kim, Gyuhyeon Nam, Geumbyeol Hwang, and Gyeongsu Chae.
\newblock Gan vocoder: Multi-resolution discriminator is all you need.
\newblock \emph{arXiv preprint arXiv:2103.05236}, 2021.

\bibitem[Zeghidour et~al.(2021)Zeghidour, Luebs, Omran, Skoglund, and
  Tagliasacchi]{zeghidour2021soundstream}
Neil Zeghidour, Alejandro Luebs, Ahmed Omran, Jan Skoglund, and Marco
  Tagliasacchi.
\newblock Soundstream: An end-to-end neural audio codec.
\newblock \emph{IEEE/ACM Transactions on Audio, Speech, and Language
  Processing}, 2021.

\end{thebibliography}
